\documentclass{comjnl}

\usepackage{amsmath}
\usepackage{float}
\restylefloat{table}
\usepackage[titletoc]{appendix}
\usepackage{amssymb}
\usepackage{graphicx}
\usepackage{verbatim,color,epsfig,cancel}
\usepackage{url}
\usepackage{enumitem}
\usepackage{amsmath }
\usepackage{todonotes}
\usepackage{balance}
\urldef{\mailsa}\path|a.pinna@diee.unica.it,matteo.orru@cs.technion.ac.il|

\begin{document}

\title[A Petri Nets Model for Blockchain Analysis]{A Petri Nets Model for Blockchain Analysis}

\author{Andrea Pinna} \author{Roberto Tonelli} 
\affiliation{%
	 Department of Electrical and Electronic Engineering (DIEE),\\ University of Cagliari, \\Piazza D'Armi, 09100 Cagliari, Italy.\\}
\email{a.pinna@diee.unica.it, roberto.tonelli@dsf.unica.it}
\author {Matteo Orr\'{u}}
\affiliation{%
	Computer Science Dept. --- Technion - IIT \\ Technion City, CS Taub Building, \\ 3200003 Haifa, Israel.\\
	}
\email{matteo.orru@cs.technion.ac.il}
\author{Michele Marchesi}%
\affiliation{	
	Department of Mathematics and Computer Science, \\ University of Cagliari, \\ Via Ospedale 72, 09124 Cagliari, Italy.
}
\email{a.pinna@diee.unica.it, roberto.tonelli@dsf.unica.it, matteo.orru@cs.technion.ac.il, marchesi@unica.it}

\shortauthors{A. Pinna, R. Tonelli, M. Orr\'{u}, M. Marchesi}

\received{00 January 2009}
\revised{00 Month 2009}


\begin{abstract}
A Blockchain is a global shared infrastructure where cryptocurrency  transactions among addresses are recorded, validated and made publicly available in a peer-to-peer network.
To date the best known and important cryptocurrency is the bitcoin. In this paper we focus on this cryptocurrency and in particular on the modeling of the Bitcoin Blockchain by using the Petri Nets formalism.
The proposed model allows us to quickly collect information about identities owning Bitcoin addresses and to recover measures and statistics on the Bitcoin network. 
By exploiting algebraic formalism, 
we reconstructed an Entities network associated to Blockchain transactions gathering together Bitcoin addresses into the single entity holding permits to manage Bitcoins held by those addresses. The model allows also to identify a set of behaviours typical of Bitcoin owners, like that of using an address only once, and to reconstruct chains for this behaviour together with the rate of firing.
Our model is highly flexible and can easily be adapted to include different features of the Bitcoin crypto-currency system.

\end{abstract}

\maketitle

\newif\ifcomment
\commentfalse

\newif\iftodo
\todofalse

\newcommand  \Andrea  	[1]  {\ifcomment \marginpar  [\emph{AP}  $\Rightarrow$]  {$\Leftarrow$  \emph{AP}}  \ap{#1} \else {} \fi}
\newcommand  \Matteo    [1]  {\ifcomment \marginpar  [\emph{MO}   $\Rightarrow$]  {$\Leftarrow$  \emph{MO}}  \mo{#1} \else {} \fi}
\newcommand  \Roberto   [1]  {\ifcomment \marginpar  [\emph{RT}   $\Rightarrow$]  {$\Leftarrow$  \emph{RT}}  \rt{#1} \else {} \fi}

\newcommand\mo[1]{\textcolor{blue}{\textsf{#1}}}
\newcommand\ap[1]{\textcolor{red}{\textsf{#1}}}
\newcommand\rt[1]{\textcolor{green}{\textsf{#1}}}

\newcommand \todom[1]{\iftodo \todo[size=\footnotesize, color=orange!60]{#1} \else {} \fi} 
\newcommand \todoim[1]{\iftodo \todo[size=\footnotesize, inline, color=orange!60]{#1} \else {} \fi} 

\makeatletter  \if@todonotes@disabled 
\newcommand{\hlfix}[2]{#1}
\else 
\newcommand{\hlfix}[2]{\texthl{#1}\todo{#2}} 
\fi 
\makeatother 

\section{Introduction}
\label{sec:Intro}
The \textit{Bitcoin electronic cash system} was conceived in the 2008  
by the scientist Satoshi Nakamoto \cite{nakamoto2009bitcoin} with the aim of producing digital coins whose control is distributed across the Internet, rather than owned by a central issuing authority, such as a government or a bank. 
It became fully operational on January 2009, when the first mining operation was completed, and since then it has constantly seen an increase in the number of users and miners. 

At the beginning, the interest in the bitcoin digital currency was purely academic, and the exchanges in bitcoins were limited to a restricted elite of people more interested in the cryptography properties  than in the real bitcoin value. 
Nowadays bitcoins are exchanged to buy and sell real goods and services as happens with traditional currencies.

The main distinctive feature introduced by the Bitcoin system is the Blockchain, that is a shared infrastructure where all bitcoins transfer are recorded. Value transfer is called \textit{transaction} and is an operation between users. 
To send and receive bitcoins, a user needs an alphanumeric code, called \textit{address}. Address represents the users' account and to each address a private key is associated.
No personal information is usually recorded in a Blockchain and for this reason Bitcoin protocol offers pseudo-anonymity. Different blockchains have been implemented so far and the technology often seems to work properly, even if most of them suffer from a lack of software engineering principles application in their development and deployment \cite{BOSE}.

To date blockchain is the technology underlying Bitcoin, but is also the technology underlying other cryptocurrencies, such as Ethereum, Litecoin and MaidSafeCoin. By analyzing this technology we can 
obtain many statistical properties of its associated cryptocurrency network, as well as the typical behavior of users, for example how users move bitcoins between their various accounts in order to preserve and reinforce their privacy.

In this paper, we introduce a novel approach, based on a Petri Net (PN) model to analyze the Blockchain. Using Petri Net we define a single useful model, a unique data structure, by which not only all main information about transactions and addresses are represented, as can be done using other approaches, but also the overall architecture and scheme of blockchain transactions are fully and natively implemented through a well known and powerful formalism.

We assume that each address corresponds to a place and each Bitcoin transaction corresponds to a transition in a Petri Net (also known as Place/Transition Net or P/T Net).
The proposed model, called ``Addresses Petri net'', allows to quickly collect information on the identities owning Bitcoin addresses and to recover measures and statistics on the Bitcoin network.  We reconstruct an Entities network associated to Block Chain transactions gathering together Bitcoin addresses into the single entity holding permits to manage Bitcoins held by those addresses.
In other words, the use of PN formalism easily allows us to construct 
first the``Addresses Petri Net'', and then the ``Entities Petri Net''.
Even if we analyzed only a few features of the bitcoin blockchain, our model perfectly 
fits blockchain behavior and features and can potentially be used to exploit the full 
behavior of this new technology and to preform statistical simulations on it. 


There a number or advantages in using PN as a model to investigate the Bitcoin transactions.

First of all, the well-defined algebraic model allows to manage straightforward algorithms to perform several structural analysis. 

Second, it allows to represent natively the Blockchain transactions, providing an 
alternative graphical representation of the Blockchain scheme. 

Finally, it opens up the possibility to perform dynamic simulations to forecast the 
future properties of the Bitcoin network. In fact the model allows the creation of higher 
level representations of the Bitcoin ledger, by grouping addresses in specific places 
and obtaining transition firing statistics.\\

The remaining of the paper is organized as follows.
In Section \ref{sec:Related} an overview of related works is reported, in Section \ref{sec:Overview} we illustrate the Bitcoin payment system, in Section  \ref{sec:PTNets} we describe our model. Specifically, Section \ref{sec:APN} illustrates the Addresses Petri Net associated to Bitcoin addresses. Section \ref{sec:EntPN} describes Entities Petri Net and the proposed algorithm used to infer from the  Entites Petri Net, the Addresses Petri Net.

In Section \ref{sec:Results} we illustrate the application of our model to the Bitcoin system and present the results.
Finally, Section \ref{sec:Discussion} presents the discussion and Section \ref{sec:Conlcusions} concludes.

\section{Related works}
\label{sec:Related}
In these last years, the unique features of Blockchain have attracted more and more researchers and several are the works that examined this shared data collection. Even if several papers focused on heuristics and algorithms in order to analyze and cluster Bitcoin addresses identifying network of users, no 
researcher focused on the analysis of the blockchain by modeling it within the framework of Petri Nets. 
This idea was carried on as the topic of the Ph.D. program of A. Pinna \cite{PHD} and some preliminary results are reported in \cite{art:Pinna}.
Consequently this section on related works will mainly focus on the works in literature which investigate blockchain technology, 
structure and properties from the point of view of dynamical networks.

Ron and Shamir \cite{Dorit:2013} analyzed and measured the Blockchain up to the block number 180,000, from January 03th, 2009 to May 13th, 2012, by using a model called \textit{transaction graph}. They analyzed the distribution of the number of transaction per address and introduced the concept of \textit{entity} as a group of addresses of the same owner. 
They ran a variant of a Union-Find graph algorithm in order to find sets of addresses belonging to the same user. First, they constructed the transaction graph, the address graph, and then constructed the contracted transaction graph and the entity graph. Thanks to this entity graph, the authors determined various statistical properties of each entity, such as the distribution of the  accumulated incoming bitcoins, the balance of bitcoins updated to May, 13th 2012, and the balance of the number of transactions per entity and per address. The authors obtained, for both the original and the clustered network (the entities 
network), some statistical properties which are typically encountered in complex networks \cite{wetsomMatteo,Newmann,santoFortunato,InfSciIvana}. 
In addition they investigated the most active entities in the system.

In \cite{Dorit:2013}, the users' common practice to move bitcoins between their various accounts (addresses) is tracked as a good practice to preserve and reinforce user's anonymity.




As a preliminary result
, the potentiality of the Petri Nets formalism for investigating users behavior has been discussed in
\cite{art:Pinna}. That work focused on the identification of ``disposable addresses'' (addresses used just once). 



Many other strategies adopted in order to preserve and reinforce users' anonymity have been analyzed in literature.
Some of these strategies improve the privacy and anonymity including mixing protocols, are discussed in CoinShuffle \cite{art:Ruffing}. 
CoinJoin and CoinParty \cite{art:Ziegeldorf} investigated the use of anonymity networks obtained by using software like TOR.
Biryukov et al. in \cite{art:Biryukov} found countermeasures to block users who access in the Bitcoin network using Tor or other similar protocols. 
Reid and Harrigan \cite{Reid} studied how an attacker could make a map of users' coins movement tracing their addresses and gathering information 
from others sources. They also focused in the topology of addresses network and transaction network, showing their properties of complex networks.
These results can be compared to those reported in \cite{matteo} for clustering other software networks. 
Androulaki \textit{et al}  \cite{androulaki} analyzed how users try to reinforce theirs anonymity in the Bitcoin system. In particular, they studied the technique of changing address and how this makes more complex the network.

Meiklejom et al. \cite{art:Meiklejom} proposed an heuristic to recognize the changing addresses method, and to keep track of potential criminal users, thanks to information extracted from the Blockchain and from other sources, such as forums. They also tried to give a name to each address. 
Kondor et al. \cite{Kondor} focus on retrieving the Blockchain transaction network, studying its features over the time. 

Recently, Lishke and Fabian \cite{art:Lischke} proposed an exploratory analysis of the Blockchain and of Bitcoin users. They studied the economy and main features of the Bitcoin cash system, but did not focus neither on the concept of "entity", nor on disposal addresses, as we do in this work. Their analysis revealed the major bitcoin businesses and markets, giving insights on the degree distribution (probability density function and complementary cumulative distribution function) of bitcoin transactions for several aggregations of time, businesses categories and country.
These distributions revealed the existence of a scale-free network, and hence that Bitcoin network follows a power law distribution although 
not over the entire period. These results can be compared to those reported in \cite{InfSciIvana} about the mechanism of power law distribution 
generation in similar technological networks and have also been replicated in this paper, 
where we found that the distributions of several investigated quantities  follow a power-law very closely (see section 5 for details).

The surge of interest regarding Bitcoin led scientists to face several other topics, in addition to the Blockchain analysis,
Cocco et al. in \cite{art:cocco2015} presented an agent-based artificial cryptocurrency market in which heterogeneous agents buy or sell cryptocurrencies, in particular Bitcoins. The model proposed is able to reproduce some of the real statistical properties of the price returns observed in the Bitcoin real market.
In \cite{art:Cocco2016} the same authors proposed an agent-based artificial cryptocurrency market in order to model the economy of the mining process. Starting from GPU's generation they reproduce some "stylized facts" found in real-time price series and some core aspects of the mining business. 

Other works focus on security and privacy issues \cite{Briere}, cryptographic problems \cite{Hernandez}, social aspects of the Bitcoin users behavior \cite{Saxena2014122,Weber:2016,Wilson} and on and economic aspects and the implication of the cryptocurrency phenomenon, see for instance works
\cite{art:Cocco2017}.


%
%

\section{The Bitcoin Cash System: an overview. }
\label{sec:Overview}
The Blockchain is a distributed and global database where all information about bitcoins' transactions are stored, but the term can also be used to denote the technology behind. 
It works as a public ledger which is composed of an ordered sequence of blocks. 
Blocks are validated and inserted into the chain and each block contains data about a variable number of validated transactions. 

Bitcoin transactions originally represented value transfer of a cryptocurrency but they can be used to transfer 
any kind of information. Each transaction is composed by an input section and an output section, which 
report a list of addresses\footnote{An alphanumeric string of 32 elements which can begin only with "1" or "4", e.g. \[1JQfVfzfxtfUb9kexSt7mHhcHxX6fyBJ5A.\];} and their associated values meaning bitcoins.


The information associated to each transaction in 
the Blockchain are characterized by: 
\begin{itemize}
	\item A list of inputs, each one containing one previous transaction; 
	\item A non empty list of outputs (possibly coinciding with some inputs);
	\item The associated amounts to each output.
\end{itemize}

\begin{figure}
	\centering
	\includegraphics[width=1\linewidth]{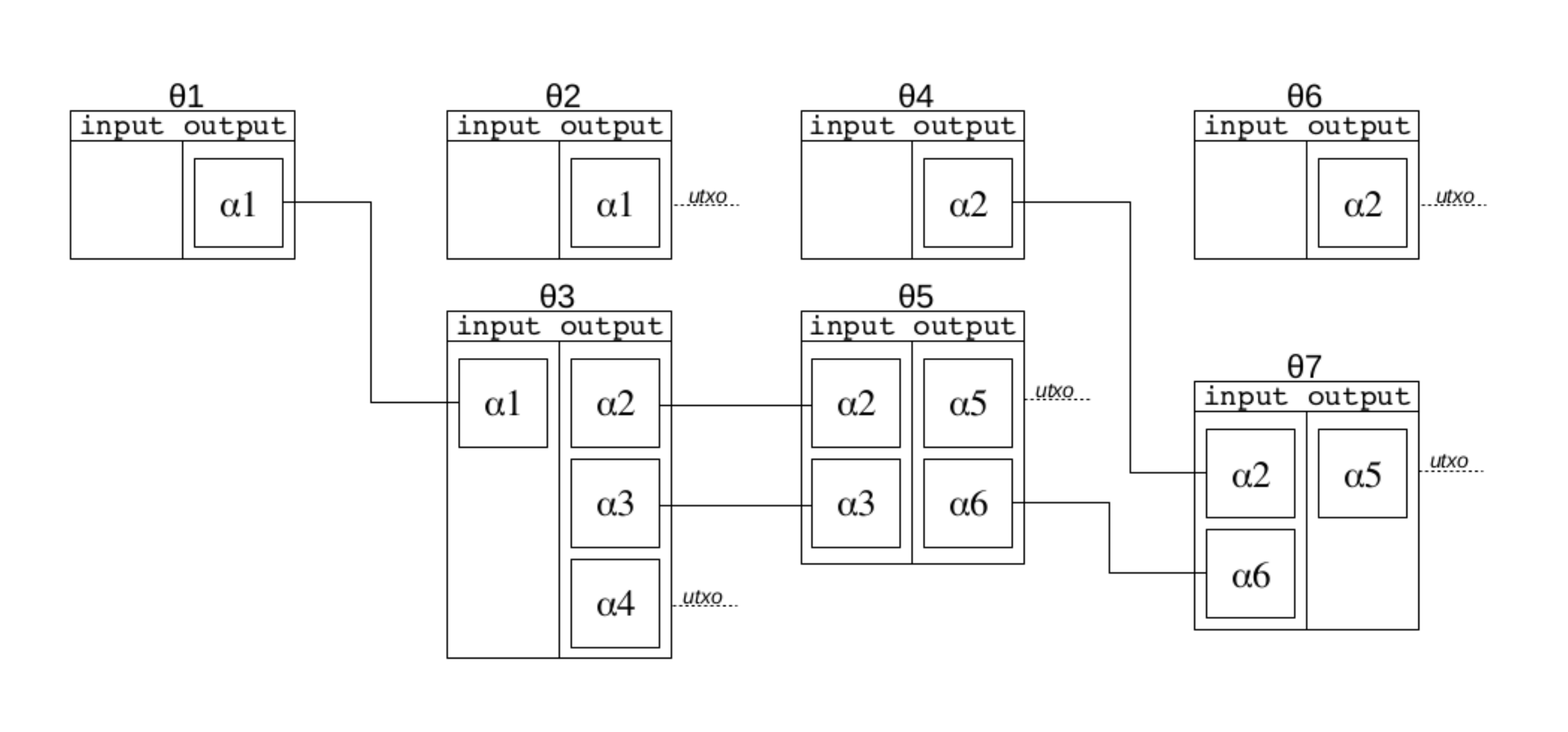}
	\caption{Simplified transaction schema.}
	\label{fig:ReteEsempio_1}
\end{figure}


Users can own one or more addresses, and address creation is costless. 
Users' anonymity is preserved since the Blockchain stores only addresses, 
and neither user names nor other identity information are required 
to create an address. 

Bitcoin clients (software which allow users to interact with the Bitcoin network)  
manage the addresses in \textit{digital wallets}. 
Wallets store both public and private keys which are used to receive and to send payments. 

Fig. \ref{fig:ReteEsempio_1} shows a simplified scheme of the interaction among transactions 
(called $\theta_i$) and addresses (called $\alpha_j$). 
In the figure seven transactions and six addresses are involved in the chains. 
The balance of bitcoins owned by users is associated to their own address, and it is equivalent 
to the total value  
of the \textit{unspent transaction outputs} (i.e., \emph{UTXO}s) 
that the address has received and not spent yet. Each square in the input section represents a spent transaction. Each square in the output section that is not connected to the input of another transaction, represents an UTXO. 
For example, the addresses $\alpha_1, \alpha_2, \alpha_4$ and $\alpha_5$ 
have one or more UTXOs, so their balance is not null. 

Each transfer of bitcoins among users implies changes on the balances associated to the respective addresses, 
similarly to what happens with a traditional bank account.
Transaction requests wait in a ``pending'' status 
in the peer-to-peer network until they are validated by miners, in order both 
to prevent frauds and to avoid double spending. 

Technical details about the network implementation can be found in \cite{nakamoto2009bitcoin}. 
Briefly, users interact with the Bitcoin network through clients which establish a Internet 
connection with some other client. 

Each client become a node of the peer-to-peer network and, potentially, each node of the Bitcoin 
system has the same importance of any other one. Nodes listen for transaction requests arriving from other nodes. 
A transaction between addresses can be accepted only if it satisfies the following constraints: 

\begin{itemize}
	\item The transaction's inputs must correspond to the outputs of previous unspent transactions (UTXO) 
	with same address and values; 
	\item The transaction's output total value must be less or equal to the total value of the inputs, 
	with a possible difference being the transaction \textit{fee}.
\end{itemize}

The validation procedure, called \textit{mining}, is carried out by \textit{miners} and 
consists in solving the (computationally hard) problem of determining an 
hash key starting with a given number of zeros (\emph{nonce}) starting from
a set of transactions requests as input. This hash key will be associated to the new \emph{validated} block.	 
In addition to transaction data, each new block contains several information such as the hash code of the previous block in the Blockchain, its height 
(its associated progressive number), and the IP address of the miner. 

Mining the blocks is a competitive task which involves all the miners in the peer-to-peer network, which try to be the first to validate the next block. 
The first miner who is able to validate a new block\footnote{who become a part of the main branch of the Blockchain, 
after eliminating the forks using a consensus rule.} receives a reward in bitcoins (presently 12.5 BTC).

A special kind of nodes in the peer-to-peer network, called \textit{full nodes}, check the new blocks, specifically their validity, also verifying that they respect the Bitcoin's core consensus rules\footnote{https://en.bitcoin.it/wiki/Full-node}.
The difficulty of this computational problem is automatically adjusted by the network, from time to time, 
in order to maintain constant, on a statistical base, the release rate of the new blocks (about a new one every ten minutes) 
and the consequent release of new Bitcoins. 

In Fig. \ref{fig:ReteEsempio_1}, we can identify the mining transactions. 
They are the transactions $\theta_1, \theta_2, \theta_4$ and $\theta_6$, 
which are the transactions having their input section empty. 
Nowadays, miners are gathered in pools to optimize the computational effort and to make constant the incoming of pool members. The whole Bitcoin system can be seen as a special typology of financial system in which, according with its technical specification, everyone can be a trader. Real time financial instruments, made possible by cloud and grid computing\cite{art:Aymerich} could aid users in that operations. 

\section{The model: the Blockchain Petri Net}
\label{sec:PTNets}
The proposed model is based on the Petri Net formalism. Using the Petri Net formalism obtained a lightweight but useful representation of the Blockchain that we call the 
Addresses Petri Net. 
Petri Net is an oriented graph, made of two types of nodes, place and transitions, where each node can be connected only with a node of the other type. 
Also the Bitcoin Blockchain can be modeled as an oriented graph, made of two types of nodes, addresses and transactions, where the latter activate transfers of tokens between the former, 
and thus can be natively modeled by using the Petri Net formalism for places and transitions, respectively.

\subsection{Petri Nets: A brief introduction}

A Petri Net \cite{art:Murata} is a formalism to describe systems based on a bipartite graph with two kind of nodes called \textit{places} and \textit{transitions}. For this reason, Petri nets are also called \textit{Place Transition nets} (\textit{P/T nets}). 
Connections between nodes are made by directed arcs. Each node can be only connected to nodes of the other type and there are two types of arcs: 
arcs ingoing into a transition, called \textit{pre-arc}, and arcs outgoing from a transition, called \textit{post-arc}. 

One of the advantages of using Petri Nets is that they are also well described by an algebraic formalism. 
The formalism provides sets to define the nodes, and matrices to describe the arcs. A Petri Net $N$ is a quadruple defined as described below. 

\begin{definition}
	\begin{equation}
	N=(P,T,Pre,Post)
	\end{equation} where
\end{definition}

\begin{itemize}

	\item \(P=\{p_1, p_2, ... , p_m\}\) is the set of $m$ places, 
	\item \(T=\{t_1, t_2, ... t_n\} \) is the set of $n$ transitions, 
	\item \( Pre : P\times T\rightarrow \mathbb{N} \) is the \emph{Pre-incidence} function
	\item \( Post : P\times T\rightarrow \mathbb{N} \) is the \emph{Post-incidence} function.
\end{itemize}

$Pre$ and $Post$ incidence functions are usually defined by mean of matrices with dimension equal to $m\times n$. 
Each element of these matrices contains the number of arcs which connect places with transitions. 
The $Pre$ matrix contains the numbers of ingoing (to transitions) arcs for each place-transition pair. 
Vice versa, each element of $Post$ matrix is the number of outgoing arcs for each place-transition pair.

Petri nets are also a powerful formalism to describe discrete event systems, as is the case of blocks generation in the Blockchain. 
To model the state of a system, a marking $M$ (i.e., a vector which defines the distribution of tokens in places) 
is needed. Transitions are aimed at modifying the marking of the system. 
Transitions absorb tokens from places connected with Pre-arcs and produce tokens for the places connected with Post-arcs, 
an operation called \textit{firing} of a transition.
Petri net and the associated initial marking form the Network system defined as $\langle N, \textbf{M}_0\rangle$, 
where $\textbf{M}_0$ is the initial marking. 
In this work we do not describe a specific state of the Blockchain so we do not need to define a marking.

\subsection{Addresses Petri Net}\label{sec:APN}

In order to obtain the Petri net algebraic representation for the Blockchain 
we provide a set theory description of the two Blockchain elements involved, 
e.g., addresses and transactions.

We denote $\mathcal{A}=\{\alpha_1,\alpha_2, ... ,\alpha_m\}$ 
the finite set of \emph{m} addresses \(\alpha\) 
registered either as inputs or outputs in the Blockchain, and with  
$\Theta=\{\theta_1,\theta_2, ... \underline{,}\theta_n\}$ the set of \emph{n} transactions $\theta$ validated by the Blockchain. 
\\

Let $N_\alpha = (P_{\alpha},T,\mathbf{PreA},\mathbf{PostA})$ be the network of addresses, where: 

\begin{itemize}
	\item \(P_{\alpha} =\{p\alpha_1, p\alpha_2, ... , p\alpha_m\}\) is the set of $m$ 
	places with each place $p\alpha$ associated to one and only one address 
	$\alpha \in \mathcal{A}$;

	\item \(T=\{t_1, t_2, ... t_n\}\) is the set of $n$ transitions where each 
	transition $t$ is associated to one and only one transaction $\theta \in \Theta$;

	\item \( \textbf{PreA} \): is the \emph{pre-incidence} matrix;

	\item \( \textbf{PostA} \): is the \emph{post-incidence} matrix.

\end{itemize}

The sets $P_{\alpha}$ and $T$ can be recovered by browsing all the 
addresses and transactions validated in the Blockchain, which are publicly available, and inserting a new place every time a new address is found, and a new transition every time a new bitcoin transaction is encountered.

In order to build the matrices \textbf{PreA} and \textbf{PostA} let us consider one transaction 
$\theta$ in the Blockchain and the associated transition $t$.
In the Blockchain, a transaction $\theta$ consists in a set of input and output addresses with the associated amounts in bitcoin.  
We denote by \(In(\theta) \subseteq \mathcal{A}\) the set of input addresses, and by \(Out(\theta) \subseteq \mathcal{A}\) the outputs set.
For each address \(\alpha \in In(\theta)\) we consider its associated place $p\alpha$ and we add a \emph{pre-arc} leaving from $p\alpha$ and arriving to the transition $t$ associated to $\theta$. At the same time, for each address \(\alpha \in Out(\theta)\) we add a \emph{post-arc} leaving from transition $t$ associated to $\theta$ and arriving to the place $p\alpha$ associated to $\alpha$.
For each couple \((p\alpha,t)\) to which a \emph{pre-arc} has been added we set \(\mathbf{PreA}(p\alpha,t)=1\), while for each couple \((p\alpha,t)\) to which a \emph{post-arc} has been added we set \(\mathbf{PostA}(p\alpha,t)=1\).

\begin{figure}
	\centering
	\includegraphics[width=0.8\linewidth]{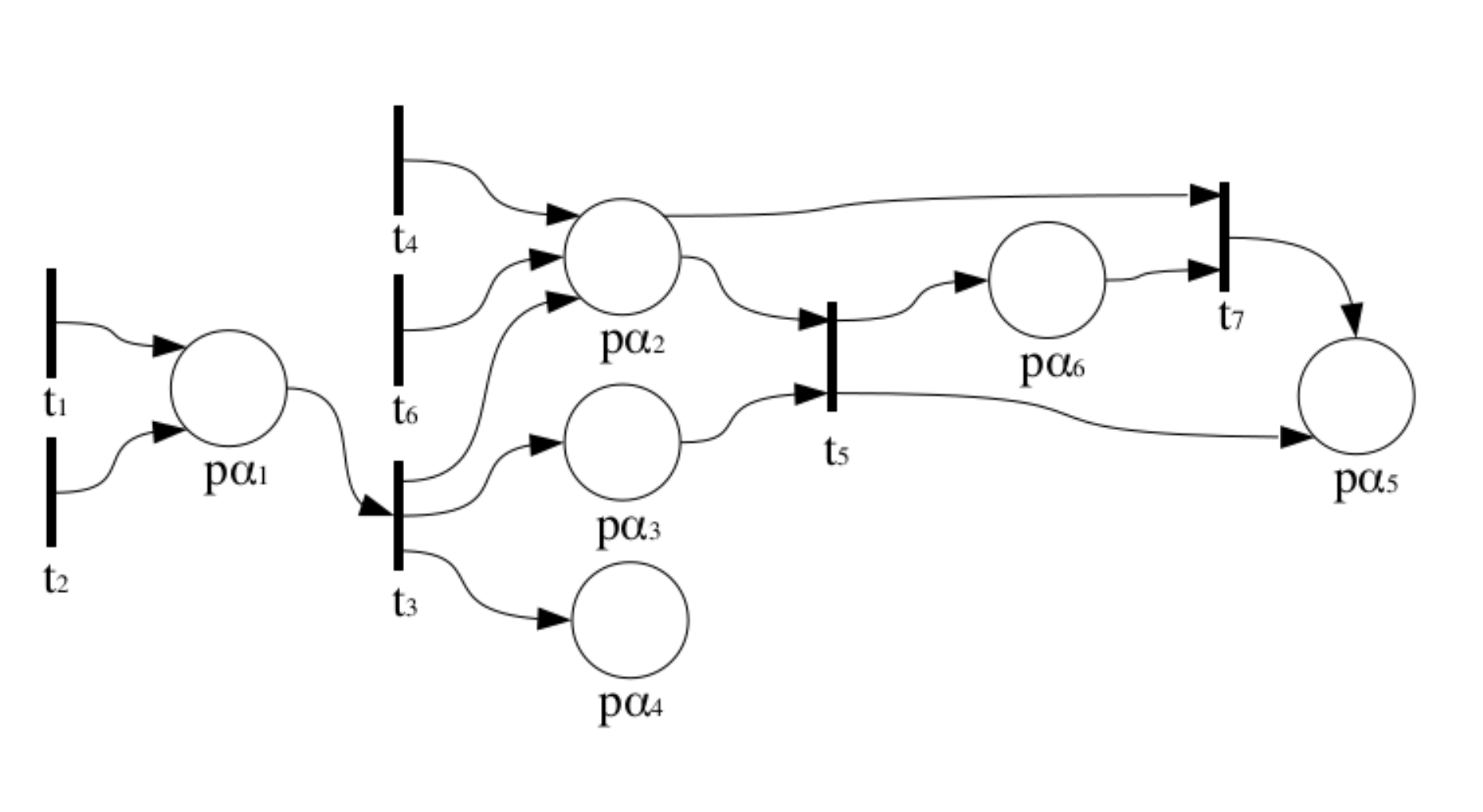}
	\caption{Addresses Petri Net equivalent to the simplified transaction chains in Fig. \ref{fig:ReteEsempio_1}}
	\label{fig:ReteEsempio_1pt}
\end{figure}

This model does not carry all the information available in the Blockchain (e.g. transactions amounts) and so it cannot completely represent Blockchain's behavior and properties. 
However, in contrast with the methodologies used in other works, in which different models were applied in order to analyze the Blockchain overloading the analysis, 
our approach natively represents the Blockchain structure and dynamics and includes into one single model and into one single data structure different features and 
properties of the Blockchain. 

Consider for instance the simplified transaction chains in Fig. \ref{fig:ReteEsempio_1}. 
There are seven transaction and six places. The equivalent Address Petri Net is composed by six places and seven transitions.
The graphical representation is shown in Fig. \ref{fig:ReteEsempio_1pt}. This Net is defined by a set of places \(P_{\alpha} = \{p\alpha_1, p\alpha_2, ... , p\alpha_6\} \), a set of transactions \(T=\{t_1, t_2, ... t_7\}\) and by the \textit{pre-} and \textit{post-incidence} matrices $\mathbf{PreA}$ and $\mathbf{PostA}$, shown in Fig. \ref{Pre-esempio1} and \ref{Post-esempio1}.

\begin{figure}
	\centering{

		\textbf{PreA}=$\begin{bmatrix} 
		0 \ & 0 \ & 1 \ & 0 \ & 0 \ & 0 \ & 0 \\
		0 \ & 0 \ & 0 \ & 0 \ & 1 \ & 0 \ & 1 \\ 
		0 \ & 0 \ & 0 \ & 0 \ & 1 \ & 0 \ & 0 \\ 
		0 \ & 0 \ & 0 \ & 0 \ & 0 \ & 0 \ & 0 \\ 
		0 \ & 0 \ & 0 \ & 0 \ & 0 \ & 0 \ & 0 \\ 
		0 \ & 0 \ & 0 \ & 0 \ & 0 \ & 0 \ & 1 \\
		\end{bmatrix}$
		$\begin{array}{c} p\alpha_1 \\p\alpha_2 \\p\alpha_3 \\p\alpha_4 \\p\alpha_5 \\p\alpha_6 \end{array}
		$\\$\begin{array}{cccccccc}\ \ & t_1 & t_2 & t_3 & t_4 & t_5 & t_6 & t_7 \end{array}$}
	\vspace{.2cm}
	\caption{Pre-incidence matrix of the Petri net for the example in Fig. \ref{fig:ReteEsempio_1}.}
	\label{Pre-esempio1}
\end{figure}

\begin{figure}
	\centering{

		\textbf{PostA}=$\begin{bmatrix} 
		1 \ & 1 \ & 0 \ & 0 \ & 0 \ & 0 \ & 0 \\
		0 \ & 0 \ & 1 \ & 1 \ & 0 \ & 1 \ & 0 \\ 
		0 \ & 0 \ & 1 \ & 0 \ & 0 \ & 0 \ & 0\\ 
		0 \ & 0 \ & 1 \ & 0 \ & 0 \ & 0 \ & 0 \\ 
		0 \ & 0 \ & 0 \ & 0 \ & 1 \ & 0 \ & 1 \\ 
		0 \ & 0 \ & 0 \ & 0 \ & 1 \ & 0 \ & 0 
		\end{bmatrix}$
		$\begin{array}{c} p\alpha_1 \\p\alpha_2 \\p\alpha_3 \\p\alpha_4 \\p\alpha_5 \\p\alpha_6 \end{array}
		$\\ $\begin{array}{cccccccc}\ \ & t_1 & t_2 & t_3 & t_4 & t_5 & t_6 & t_7 \end{array}$}
	\vspace{.2cm}
	\caption{Post-incidence matrix of the Petri Net for the example in Fig. \ref{fig:ReteEsempio_1}.}
	\label{Post-esempio1}
\end{figure}

These matrices can be straightforwardly used to perform several analysis of the network. 
For example, we can compute the difference between \textit{post} and \textit{pre-incidence} matrices and consider one of its row. 
The number of not null elements in such row is equal to the number of UTXO contained in the address related to the place corresponding to the row. 
This number must be greater than or equal to zero, and if it is equal to zero the balance of the associated address is null. 

In addition, we can easily compute the number of times that an address appears as input in a transaction. 
In fact, all the not-zero elements of the row \emph{i} of the matrix \( \mathbf{PreA} \) provide the number 
of times the address $\alpha$ corresponding to the place \(p\alpha=i\) has been the input of a transaction.


As other example we consider the case of different transactions occurring in different moments which share the same input set and the same output set. 
Using our model, these transactions can be represented with only one transition, which is characterized by a firing clock. 
This feature, along with the creation of the entities net, can be useful to enable a dynamical and high level analysis of the 
Bitcoin system. We will show in the following that our model allows to easily detect such sets of transactions.

\subsection{Entities Petri Net and algorithm to manage them}\label{sec:EntPN}
It is quite common for Bitcoin users to hold more than one address in order to manage bitcoin exchanges and anonymity more easily.
As in \cite{Dorit:2013} we define an \textit{entity} as the person, the organization, the group of people, or the firm that hold 
the control of the bitcoins associated to a set of addresses. 
All addresses appearing in an input section of a single transaction must be owned by the same entity. 
This is because, in order to activate the bitcoin transfers from those addresses, the same entity must hold 
all the private keys of all corresponding wallets 


In order to build the Entities Petri Net \(N_{\epsilon}\) we associated each entity 
to a collection of addresses, associating places $p\epsilon \in P_{\epsilon}$ in \(N_{\epsilon}\) 
to a set of places $p\alpha$ of \(N_{\alpha}\).
We denote by \(E=\{\epsilon_1,\epsilon_2, ... ,\epsilon_k\}\) 
the set of \textit{entities} where each \textit{entity} \(\epsilon \in E\) is 
a finite set of addresses such that $\epsilon\subseteq \mathcal{A}$.

The matrix \(\mathbf{PreA}\) has $m$ rows, 
one for each place, and $n$ columns, one for each transition. 
Given a transition $t$ we consider the array \(\mathbf{PreA}(\cdot,t)\) 
which is the column of \(\mathbf{PreA}\) with index $t$. 
Its non zero elements correspond to places $p\alpha$ 
with \(\mathbf{PreA}(p\alpha,t)=1\), namely places with outgoing arcs \emph{pre-arc} towards transition \(t\). 
These places \(p\alpha\) correspond to input addresses \( \alpha \in In(\theta)\), for the transaction $\theta$ corresponding to transition $t$. 
As a consequence, all these places belong to one single entity \( \epsilon \in E\).

It is also possible that a given address appears in two or more input sections, together with other addresses. In this case, the entity must be composed by all the addresses in these input sections.

To build the Entities Petri Net, $E$, we applied the following algorithm.

\begin{figure}
Let be $T^*=T$ the set of unexplored transitions and $E=\emptyset$ the set of entities.
\begin{itemize}
	\item while $T^*\neq \emptyset  $
	\begin{enumerate} 
		\item take a $t : t \in T^{*}$ and remove this form $T^{*}$
		\item let $e=\emptyset$
		\item for all $i: \mathbf{PreA}(p_i,t) \neq 0$ do
		$e=e \cup \{p_i\}$

		\item let $e^*=e$ the set of unexplored places
		\item while $e^*\neq \emptyset$ 
		\begin{enumerate}
			\item take a place $p \in e^*$
			\item let $T'= \emptyset $
			\item for all $j:\mathbf{PreA}(p,t_j) \neq 0$ do $T'= T' \cup \{t'\} $

			\item for all $t' \in T'$
			\begin{enumerate}
				\item let $e_{new} = \emptyset$
				\item for all $h:\mathbf{PreA}(p_h,t') \neq 0$ do  $e_{new} = e_{new} \cup \{p_h\}$  
				endfor
				\item $e = e \cup e_{new}$ and $e^*= e^* \cup e_{new}$
				\item $e^*= e^* \setminus p$
				\item $T^{*}=T^{*} \setminus t'$
			\end{enumerate}
			endfor
		\end{enumerate}
		endwhile
		\item $E=E\cup e$
	\end{enumerate}
	\item endwhile
\end{itemize}
\caption{Algorithm used to compute the set $E$ of entities.}
\label{fig:algorithm}
\end{figure}

We denote \textit{unexplored place}, every place which is an element of the current entity, 
but is not yet processed. In fact, in order to find other places to be inserted into the current 
entity $e$, each unexplored place must be processed as in step 5. In this step, all the other 
places $p_h$ element of $e$ are found. 

Each $e \in E$ is a set of places of the Addresses Petri Net or, equivalently, is the representation of a set of addresses that compose an entity. 


The algorithm creates the set $E$ of entities. 
The correctness of the algorithm can be discussed analyzing the two requirements: 
the finite number of iterations and the correctness of the solution.
Firstly, the number of iterations is limited by the number of transitions. 
In fact, the set of unexplored transitions will be emptied every time a transition will be examined. 
In particular, both in step 1 and in step 5.d.v. 
a transition is removed from $T^{*}$.
Regarding the second point, because place determination occurs by evaluating the \textit{pre-arcs} connected to each transition, 
entities are correctly created and populated. Furthermore, it is possible to check that the resulting entities form mutually 
disjoint sets and that the result of the entities' union contains all the places of the Addresses Petri net. 

We can define $N_{\epsilon}$, the Entity Petri Net, as $N_{\epsilon} = (P_{\epsilon},T,\mathbf{PreE},\mathbf{PostE}) $,
where $P_{\epsilon}$ is the set of places that are associated one to one with elements of the entities set $E$.

The definition includes the set $T$ of transitions. This is the same that we have in the Addresses Petri Net.


In order to compute $ \mathbf{PreE}$ and $\mathbf{PostE}$ rows, we take every entity $e \in E$. Given an entity $e$, we first extract from $ \mathbf{PreA}$ and then from $\mathbf{PostE}$ the rows 
corresponding to every place $p_{\alpha} \in e$. Then, for each matrix, we sum these rows together. 
In this way, we obtain one new row for both $\mathbf{PreE}$ and  $\mathbf{PostE}$, corresponding to the entity $e$.
\\



For instance, looking at the Address Petri Net in Fig. \ref{fig:ReteEsempio_1pt} and at the $\mathbf{PreA}$ matrix, we recognize that places $p\alpha_2, p\alpha_3$ and $p\alpha_6$ can be joined to an entity, 
and that hence their related addresses $\alpha_2, \alpha_3, \alpha_6$ are owned by the same person. 
In total four entities are recognized as described in Table \ref{tab:EntityEsempio}.

\begin{table}
	\begin{center}
		\begin{tabular}{|c|c|}
			\hline
			Entity in $E$ & Places  \\ \hline 
			$e_1$  & $\{p\alpha_1\}$  \\ 
			$e_2$  & $\{p\alpha_2, p\alpha_3, p\alpha_6\}$ \\ 
			$e_3$  & $\{p\alpha_4\}$ \\ 
			$e_4$  & $\{p\alpha_5\}$ \\ 
			\hline
	\end{tabular}
		\vspace{.2cm}
	\end{center}
	\caption{Entity in the Entities Petri Net of the simplified transaction chains in Fig. \ref{fig:ReteEsempio_1}. }
	\label{tab:EntityEsempio}
\end{table}

\begin{figure}
\centering
\includegraphics[width=0.7\linewidth]{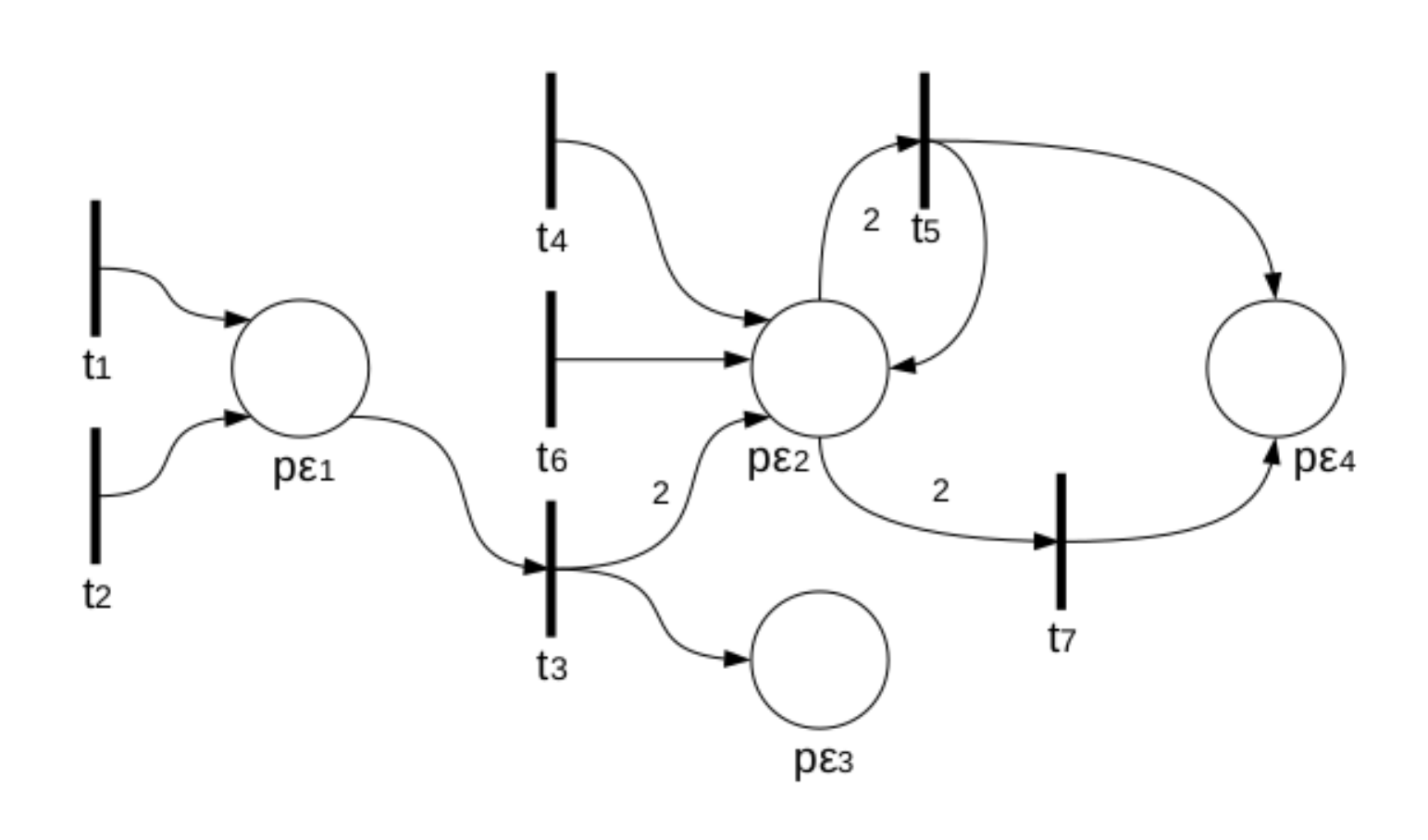}
\caption{The Entities Petri Net of the simplified transaction chains in Fig.\ref{fig:ReteEsempio_1}}

\label{fig:reteesemopioept}
\end{figure}

To each entity, a place $p\epsilon \in P_\epsilon$ is then associated. In the following tables, $\mathbf{PreE}$ and $\mathbf{PostE}$ of the example resulting Entities Petri Net are shown in Fig. \ref{Pre-E-esempio1} and \ref{Post-E-esempio1}. In Fig \ref{fig:reteesemopioept}, the graphic representation of the Entities Petri net is shown.

\begin{figure}
	\centering{

		\textbf{PreE}=$\begin{bmatrix} 
		0 \ & 0 \ & 1 \ & 0 \ & 0 \ & 0 \ & 0 \\
		0 \ & 0 \ & 0 \ & 0 \ & 2 \ & 0 \ & 2 \\
		0 \ & 0 \ & 0 \ & 0 \ & 0 \ & 0 \ & 0 \\ 
		0 \ & 0 \ & 0 \ & 0 \ & 0 \ & 0 \ & 0 
		\end{bmatrix}$
		$\begin{array}{c} p\epsilon_1 \\p\epsilon_2 \\p\epsilon_3 \\p\epsilon_4 \end{array}
		$\\$\begin{array}{cccccccc}\ \ & t_1 & t_2 & t_3 & t_4 & t_5 & t_6 & t_7 \end{array}$}
	\vspace{.2cm}
	\caption{Pre-incidence matrix of the Entities Petri Net for the simplified transaction chains in Fig. \ref{fig:ReteEsempio_1}.}
	\label{Pre-E-esempio1}
\end{figure}

\begin{figure}
	\centering{

		\textbf{PostE}=$\begin{bmatrix} 
		1 \ & 1 \ & 0 \ & 0 \ & 0 \ & 0 \ & 0 \\
		0 \ & 0 \ & 2 \ & 1 \ & 1 \ & 1 \ & 0 \\ 
		0 \ & 0 \ & 1 \ & 0 \ & 0 \ & 0 \ & 0 \\ 
		0 \ & 0 \ & 0 \ & 0 \ & 1 \ & 0 \ & 1  

		\end{bmatrix}$
		$\begin{array}{c} p\epsilon_1 \\p\epsilon_2 \\p\epsilon_3 \\p\epsilon_4 \end{array}
		$\\ $\begin{array}{cccccccc}\ \ & t_1 & t_2 & t_3 & t_4 & t_5 & t_6 & t_7 \end{array}$}
	\vspace{.2cm}
	\caption{Post-incidence matrix of the Entities Petri Net for the simplified transaction chains in Fig. \ref{fig:ReteEsempio_1}.}
	\label{Post-E-esempio1}
\end{figure}

\section{Results}
\label{sec:Results}
Blockchain can be explored mainly using two approaches. The first consists in downloading all binary data from the peer-to-peer network, and in identifying transactions, addresses and other information by using protocol instructions.  The second one consists in exploring specific websites where the decoded Blockchain is shown, and application interfaces or other utilities, are provided to explore it. 
We followed the second approach and downloaded blocks as formatted JSON files from the  website \textit{blockchain.info}. 

We parsed the first 180,000 blocks in the Blockchain, corresponding to 
a period of about three and half years, from January 2009 to March 2012, in order to compare our results with those in \cite{Dorit:2013}. 
\begin{figure*}
	\centering
	\includegraphics[width=0.8\linewidth]{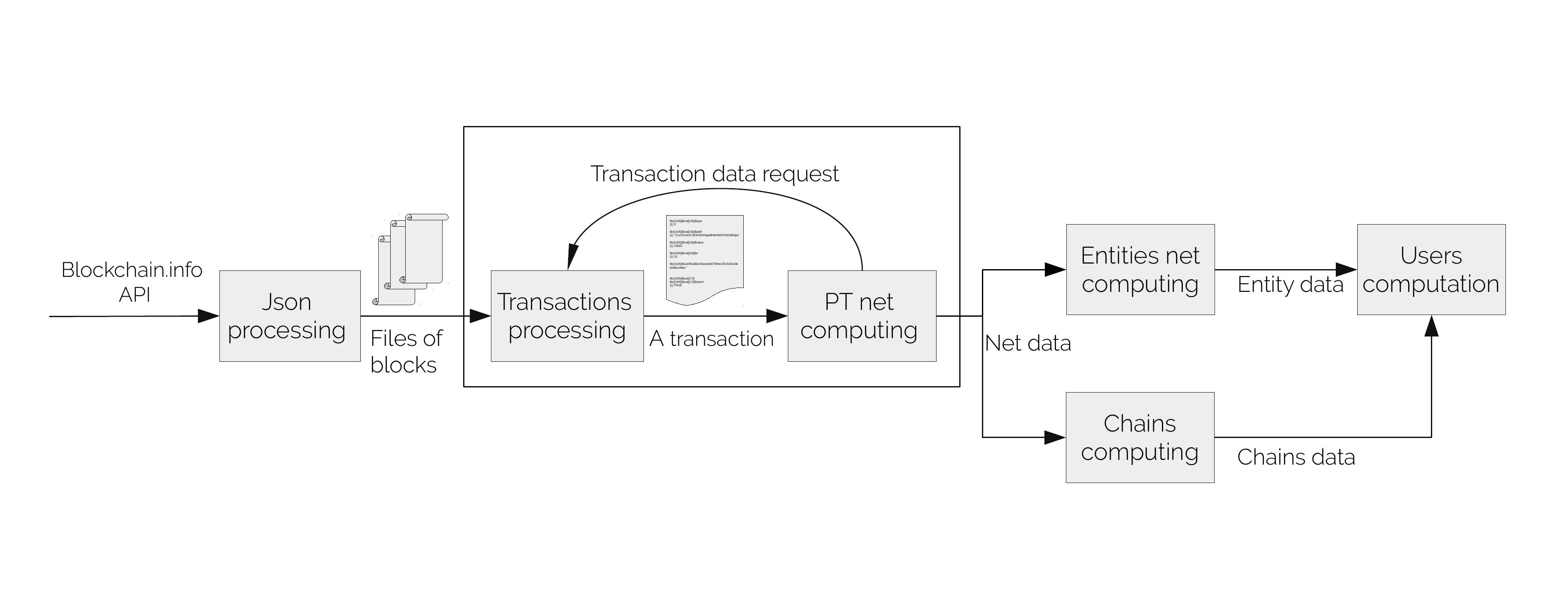}
	\caption{Diagram of the data processing path for the study of Blockchain}
	\label{fig:diagram}
\end{figure*}

The data processing performed in this work is carried on in steps as shown in Fig. \ref{fig:diagram}. All implementations are made with R language and RStudio IDE. 

%
The analyzed portion of Blockchain was processed without specific hardware resources and processing 
time to elaborate the first 180,000 blocks has been about 250 hours long. The average time required to compute a block is 5 seconds. The single block computation time depends on the number of addresses contained in it, considering every address in input or in output of a transaction. The procedure requires about ten seconds to elaborate eight addresses and does not increase significantly even when matrices become larger.  

	
The situation has quite changed for blocks validated in subsequent periods. 
Currently a block contains about three thousand addresses and the time to compute it using our Petri Nets modeling is about six minutes long. 
Generally the time to elaborate an addresses is larger if the address has not yet been found and the search algorithm must add it into the matrices. 
The downloaded JSON files, elaborated and saved in a R structure, occupies 2.8GB and after the elaboration the addresses Petri net occupies about 800MB in RAM. 
Saving corresponding data in a Rdata file, it occupies about 40MB.
\\

\subsection{Results of the Addresses Petri net}

We found 3,730,480 different addresses and 3,142,019 transactions, which in our model correspond to the number of rows and columns of matrices $\mathbf{PreA}$ or $\mathbf{PostA}$. 
We associated the addresses to the corresponding places in the set $P_{\alpha}$ in the Petri Net $N\alpha$. 
From the analysis of the matrices $\mathbf{PreA}$ and $\mathbf{PostA}$, simply counting the non zero elements, we found 4.575.888 \textit{pre-arcs} and 7.352.494 
\textit{post-arcs} in total. 
The number of non zero elements $L(i)$ on the corresponding row of 
$\mathbf{PreA}(p\alpha_i, \cdot)$ represents the 
number of transitions occurring from the place $p\alpha_i$ through a \textit{pre-arc}.
The number of non zero elements $L(i)$ in the row $\mathbf{PostA}(p\alpha_i, \cdot)$ represents the number of transitions connected to the place $p\alpha_i$ through a \textit{post-arc}.
Using this formalism our model easily takes into account the total number of bitcoin transactions in input and output of each address. 

\begin{figure}[htbp]
	\centering
	\begin{minipage}[c]{.45\textwidth}
		\includegraphics[width=7.5cm]{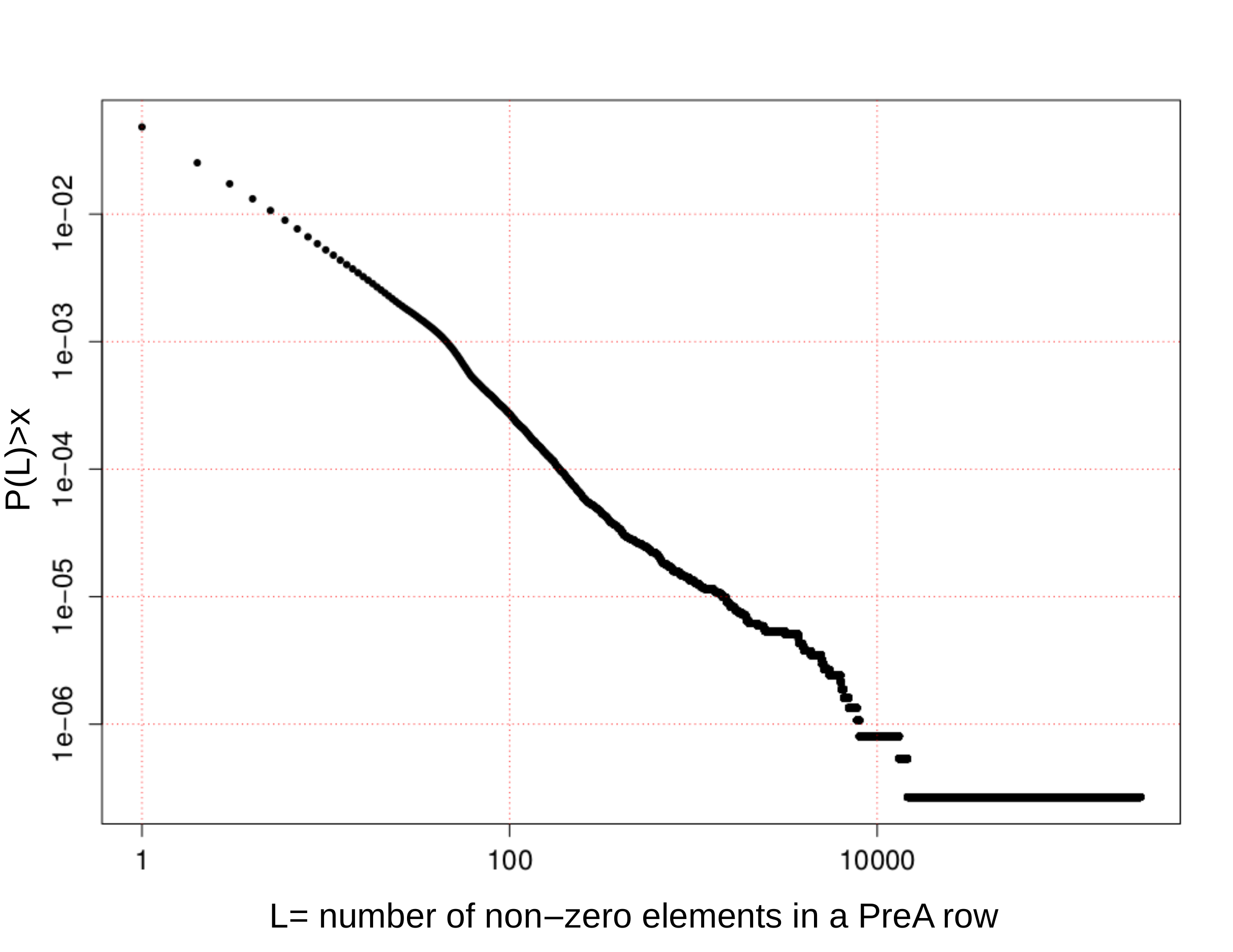} 
		\caption{CCDF of the length L for $PreA$. }
		\label{fig_preA_CCDF}
	\end{minipage}%
	\hspace{10mm}%
	\begin{minipage}[c]{.45\textwidth}

		\includegraphics[width=7.5cm]{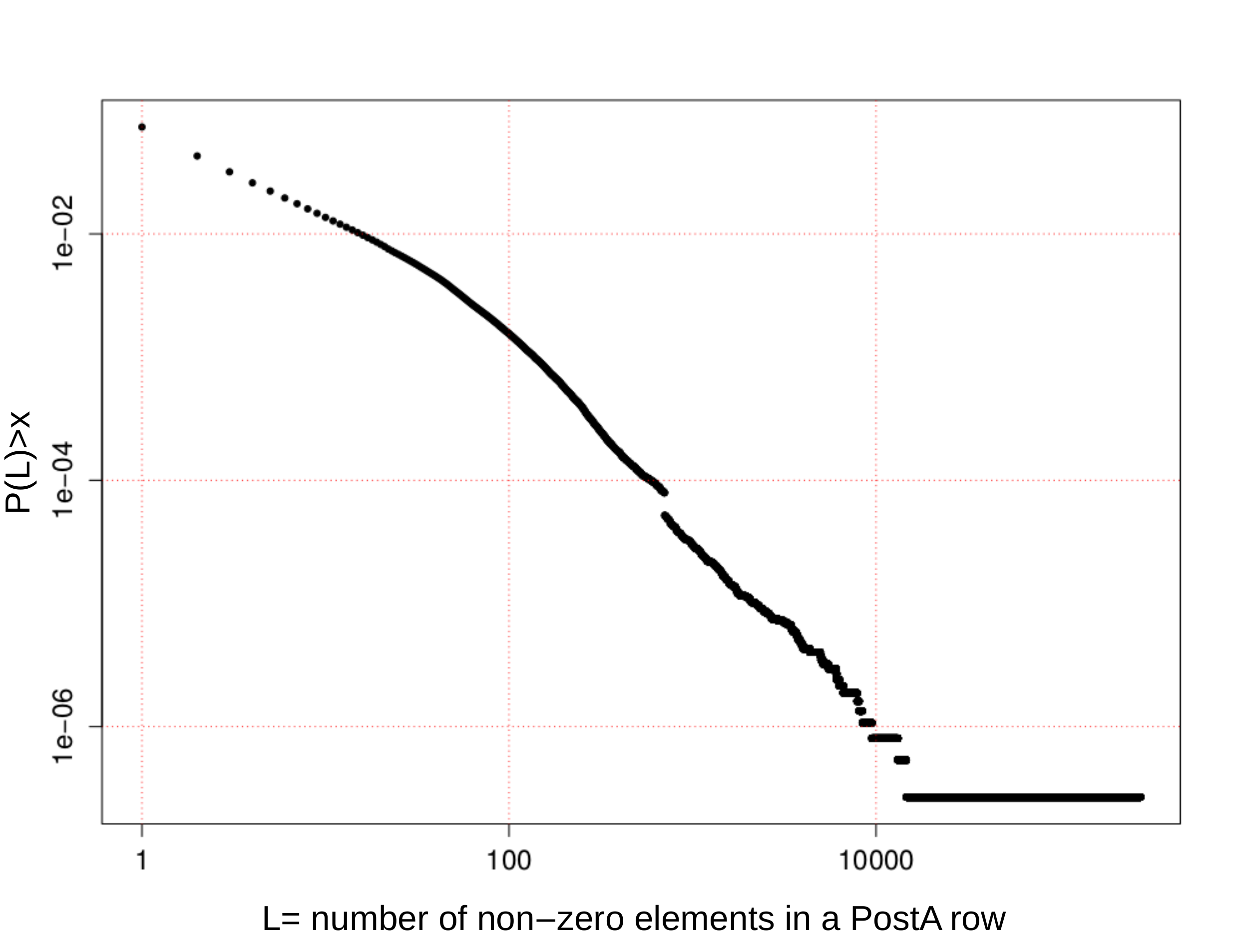} 
		\caption{CCDF of the length L for $PostA$. }
		\label{fig_postA_CCDF}	\end{minipage}%
\end{figure}

Figures \ref{fig_preA_CCDF} and \ref{fig_postA_CCDF} report the Complementary Cumulative Distribution Functions (CCDF) 
defined as the probability $P$ that $P(L)>x$, where $L$ is defined as the number of non-zero elements in the matrices $\mathbf{PreA}$ and $\mathbf{PostA}$ respectively.   
\\

The figures show an uneven distribution of $in$ and $out$ transactions among addresses so that there are many 
addresses with few transactions and relatively few addresses with many transactions, displaying a typical 
power-law distribution. Such distribution has been straightforwardly recovered using the Petri Nets formalism.

In table \ref{tab_10addressPreA} we report the ten most used addresses, 
found summing up the number of non zero elements in $\mathbf{PreA}$ rows
to that of non zero elements in $\mathbf{PostA}$ rows.

Our analysis identifies also 609,295 addresses with only zeros in 
$\mathbf{PreA}$ rows, and at least one non null element in $\mathbf{PostA}$ rows, namely 609,295 addresses never used to spend (until the 180,000 block), but 
only to accumulate. Part of them are still unused up to today.
Table \ref{tab_imbalance_10addressPreA} reports the first ten ranked by 
the number of \textit{post-arcs}, namely the number of incoming transactions, 
and shows their current balances, as checked from 
blockchain.info: four of these (row 1,7,8 and 9 in the tab) are never used again, and can be called \textit{dormant}. 
Their balance can be quite high, since they've been used like sort of bitcoin deposits. 

%
%

\begin{table*}[ht]
	\begin{center}
		\begin{tabular}{cccc} 

				\hline
				Address &  L pre  & L post & tags \\ \hline 
				1VayNert3x1KzbpzMGt2qdqrAThiRovi8  & 270,204  & 275,398 & deepbit.net \\ 
				1dice8EMZmqKvrGE4Qc9bUFf9PX3xaYDp & 14,606  & 14,605 & SatoshiDICE 48$\%$\\ 
				1dice97ECuByXAvqXpaYzSaQuPVvrtmz6  & 13,137  & 13,124 & SatoshiDICE 50$\%$\\ 
				159FTr7Gjs2Qbj4Q5q29cvmchhqymQA7of & 8,016  & 8,425 & - spammer ? - \\ 
				1CDysWzQ5Z4hMLhsj4AKAEFwrgXRC8DqRN & 6,382  & 9,501 & Instawallet\\ 
				1E29AKE7Lh1xW4ujHotoT4JVDaDdRPJnWu & 7,761  &  8,079 & - unknow -\\ 
				15VjRaDX9zpbA8LVnbrCAFzrVzN7ixHNsC & 6,999  & 7,888 & faucet donation \\ 
				15ArtCgi3wmpQAAfYx4riaFmo4prJA4VsK & 6,578  & 6,622 & faucet donation \\ 
				1dice9wcMu5hLF4g81u8nioL5mmSHTApw  & 6,318   & 6,306 & SatoshiDICE 73$\%$  \\ 
				1Bw1hpkUrTKRmrwJBGdZTenoFeX63zrq33 & 5,498  & 5,498 & - unknow - \\ \hline

			\end{tabular}
			\vspace{.2cm}
		\end{center}
		\caption{Summary of first 10 most used addresses}
		\label{tab_10addressPreA}
	\end{table*}

	\begin{table*}[ht]
		\begin{center}

			\begin{tabular}{cccc} 
					\hline
					Address &  L post & current balance BTC \\ \hline 
					15S1TFTosxrgZxkqJR2n1AFJ22ZJE2rTCk  & 3,853  &  120.85215349  \\ 
					1PtnGiNvhAKbuUQ6nZ7nF3CDKCKGfeMsCX & 1,199  &  0\\ 
					129FTwWoi5H5ujasMZ6M6VjJzBJfsXVQGw  & 1,138  &  0.78425567 \\ 
					1FN9kKsZA9XttrAwuDDgsXjs6CXUR2fzmt & 1,111 &  0  \\ 
					1DYvtKtZ2Ay9vTjzjb9BiRauMgXdjRDaD &973  &  14.5601 \\ 
					1STRonGxnFTeJiA7pgyneKknR29AwBM77 & 949  &   	1.79274504 \\ 
					1Q3nqtUzBp6jw7opi674Pyfgu4MUmVRdrk & 861  & 16.31551365 \\ 
					1Hh3eNNqR8MajEtDfvUF3hoxgf8CuUXVwY & 819  &  257.32881319 \\ 
					14sx4sFdUE9YDpJ9XbD6xAUEKPKvc8QHq2  & 811   & 59.56546509  \\ 
					17igtzSD39ZAapsut2DQTTKFyqSp7CToMq &  809  &   	0  \\ \hline

				\end{tabular}
				\vspace{.2cm}
			\end{center}
			\caption{Summary of first 10 most imbalanced addresses}
			\label{tab_imbalance_10addressPreA}
	\end{table*}

In order to analyze users practices for preserving anonymity, we focused on recognizing chains of transaction where \textit{disposable addresses} are involved.  
As already mentioned, a disposable address is an address used only two times: one time to receive bitcoins and one time to give away all these bitcoins.
Transactions which involve disposable addresses have only a disposable address in the input section and one disposable address in the output section, together with other addresses.
Usually, in the output section only two addresses in total are present. This practice is commonly used and can be performed automatically. Users who adopt it, usually give rise a long chain of transactions without waiting for confirmation. 

With our model, \textit{disposable addresses} and their chains can be easily traced analyzing $\mathbf{PreA}$ and $\mathbf{PostA}$ matrices. 
To identify the involved transactions we identified the correspondent transitions having only a pre-arc and two post-arc in the Addresses Petri Net. These are transitions that correspond to columns of $\mathbf{PreA}$ having only one non zero element and to columns of $\mathbf{PostA}$ having two non zero elements but in different rows. 

Disposable addresses are likewise identified through the correspondents places. 

Give a place, the correspondent row of the Pre and Post matrices must have one and only one non zero element. 

Under these conditions, it is possible that an Address Petri net transition becomes a cyclic transition in the Entities Petri net. In fact, in this case, output addresses and input addresses of the associated transaction are ascribed to the same entity. 
This indicated that the owner of the transaction wanted to move coins between addresses in his possession.\\

\begin{figure}
	\centering

	\includegraphics[width=7.5cm]{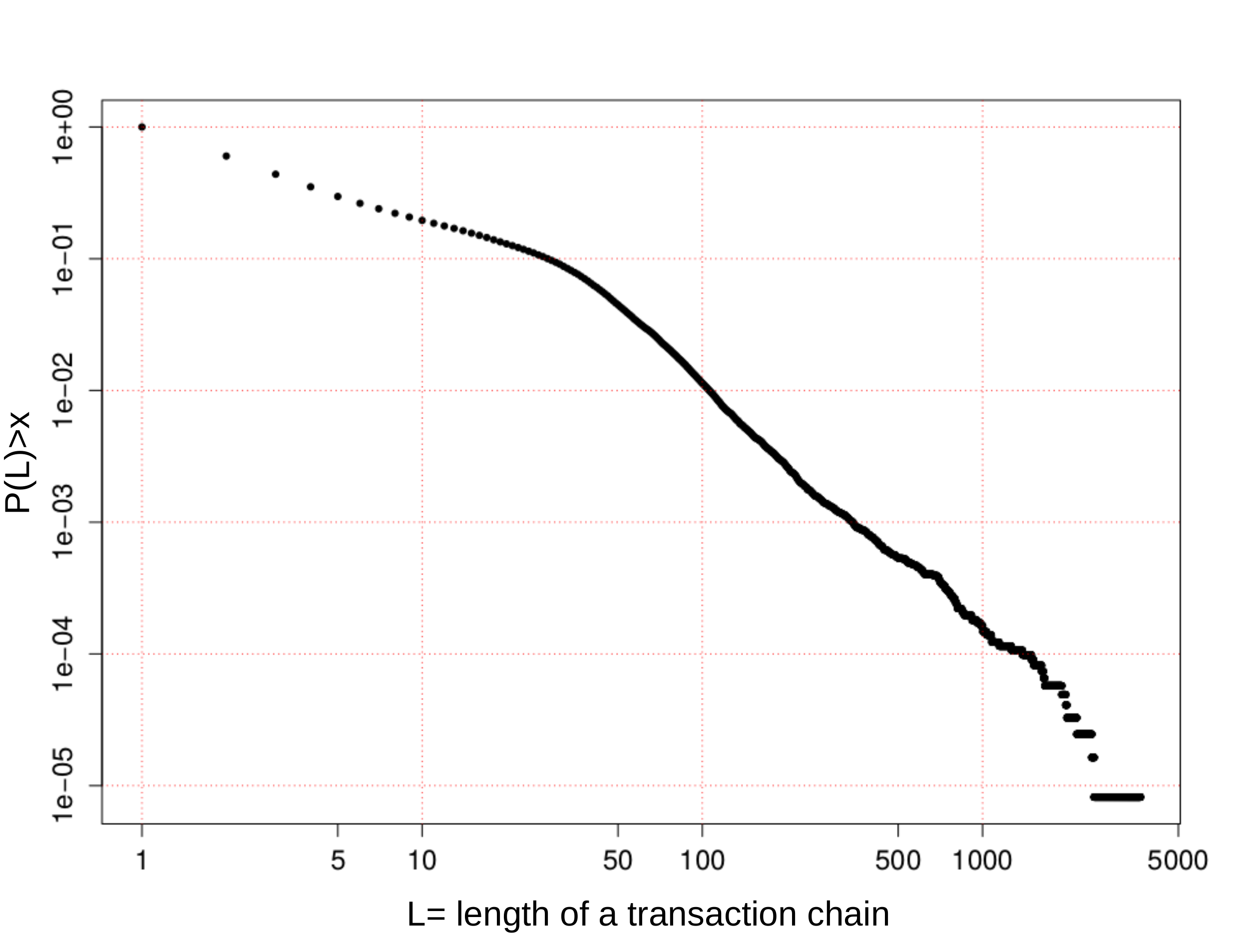}
	\caption{CCDF of the distribution of the length of the chains.}
	\label{fig_chains_CCDF}

	\hspace{10mm}%

\end{figure}

The algorithm developed to build the chains of this kind of transitions is described in detail in Appendix A. 
Applying this algorithm we found 122,155 disposable address chains, involving 1,350,010 different addresses and transactions. 
Figure \ref{fig_chains_CCDF} reports the CCDF of the chains lengths, showing 
that these are unevenly distributed, with the longest chain counting 3,658 transactions. 

We also counted how many times users repeat the same transaction in terms of the same set of addresses in input section and the same set of addresses in output section. In our model identifying these repetitions is trivial. When two or more transactions involve identical sets of addresses in input and output, the corresponding transitions are connected with the same places both in pre and in post matrices. 

\begin{figure}
	\centering
	\includegraphics[width=7.5cm]{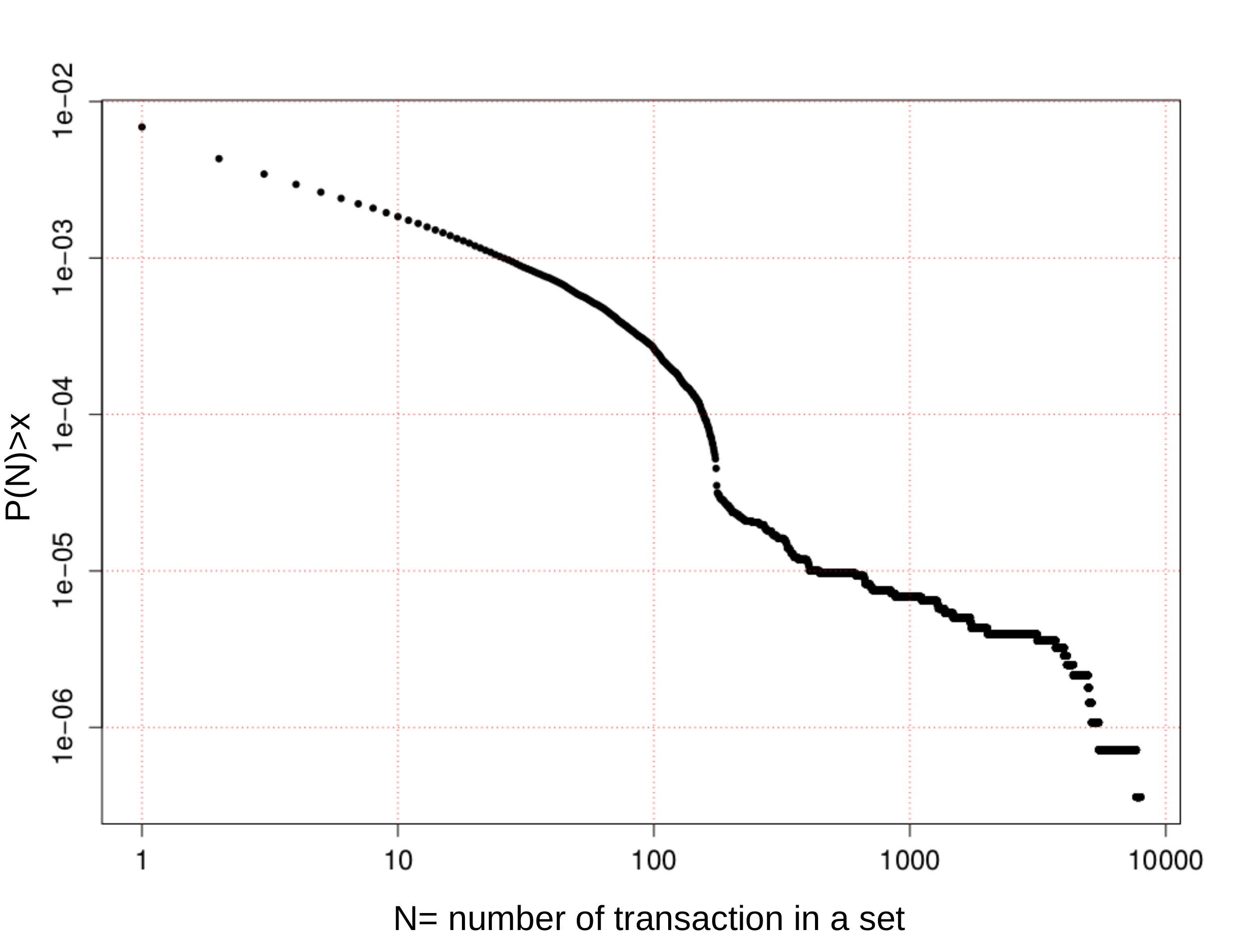} 
	\caption{CCDF of the size L of grouped transaction set for the address net }
	\label{fig_transA_CCDF}
	\hspace{10mm}%
\end{figure}


In fact, taking the matrix $\mathbf{PA}= (\mathbf{PreA}, \mathbf{PostA})$, in which the two matrices are concatenated in column, 
for each column $t_j$ it is possible to check the existence of other identical columns. 
\\

%
%
%
%

We found that about $11\%$ of transactions are a repetition of another one. These represent repeated transfer of bitcoins from one group of addresses to another group of addresses where the two groups are always the same, revealing steady fluxes of bitcoins. Figure \ref{fig_transA_CCDF} 
reports the CCDF for the sizes of these groups of repeated transactions.

\subsection{Results of the Entities Petri Net}
\begin{figure}

	\centering
	\includegraphics[width=7.5cm]{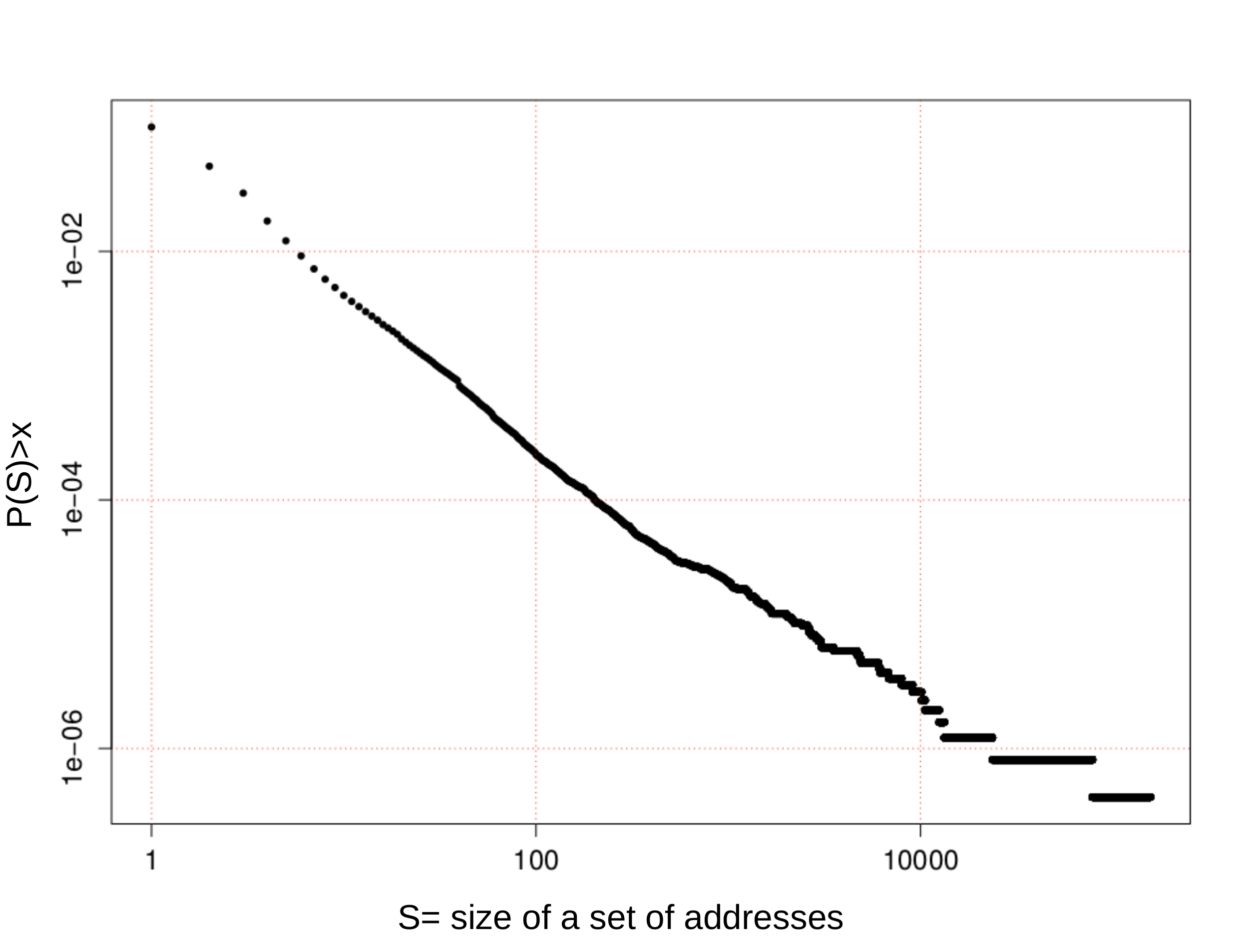}
	\caption{CCDF of the distribution of addresses across entities.}
	\label{fig_entity_CCDF}

\end{figure}

The reducing algorithm discussed in section \ref{sec:EntPN}  is applied to the Addresses Petri Net in order to recover the corresponding Entities Petri Net. Among the owners, we found that 2,461,010 entities hold all the 3,730,480 addresses, and the distribution of addresses among entities is highly not uniform. 
Figure \ref{fig_entity_CCDF} shows that also such distribution follows a power-law very closely. 
This means that there are many entities holding a single address 
but also a few entities controlling very many addresses, and thus able to control a great fraction of the bitcoins flux transactions.  

\begin{figure}[htbp]
	\centering
	\begin{minipage}[c]{.45\textwidth}

		\includegraphics[width=7.5cm]{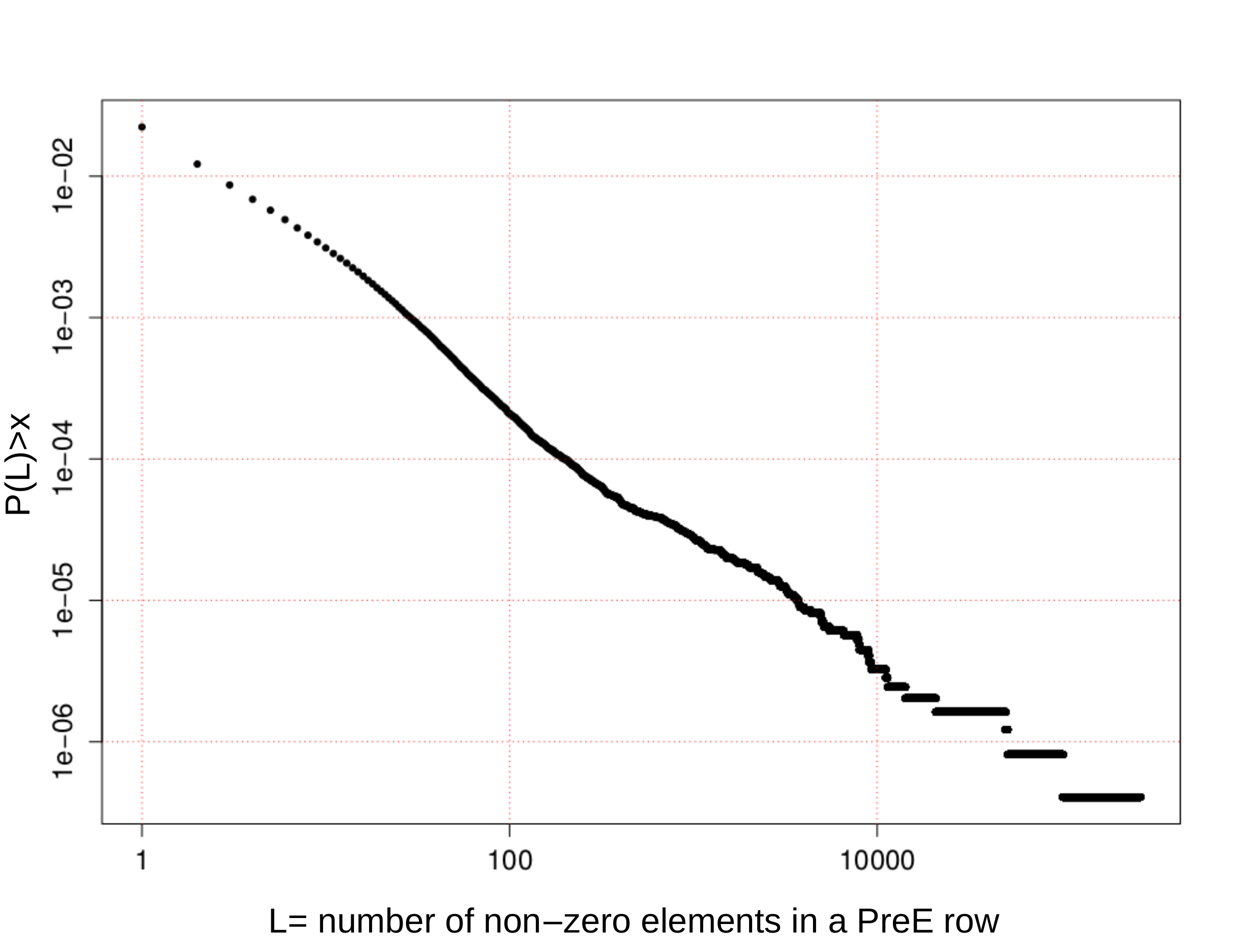} 
		\caption{CCDF of the lenght L for $PreE$. }
		\label{fig_preE_CCDF}
	\end{minipage}
	\hspace{10mm}%
	\begin{minipage}[c]{.45\textwidth}
		\includegraphics[width=7.5cm]{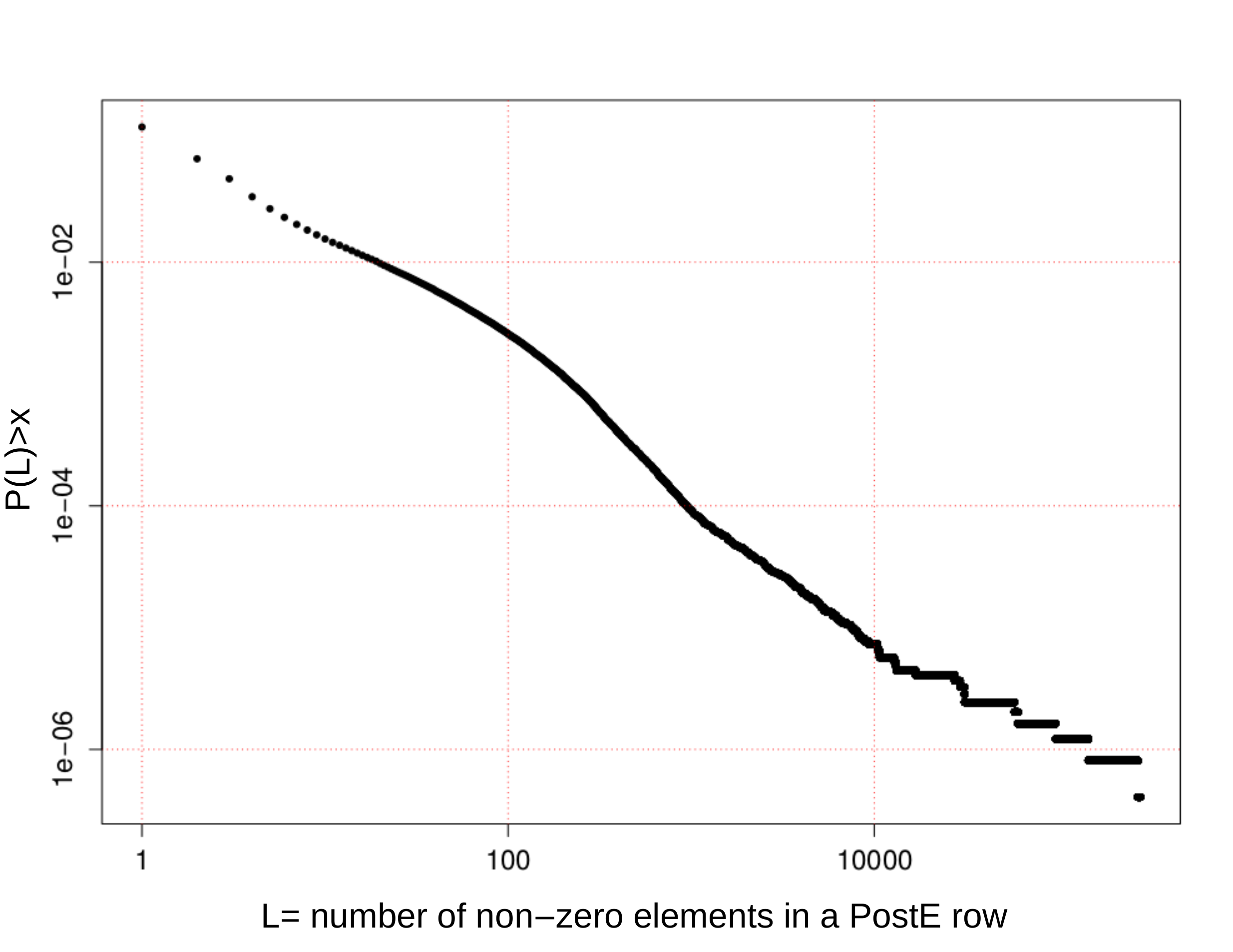} 
		\caption{CCDF of the lenght L for $PostE$.}
		\label{fig_postE_CCDF}
	\end{minipage}
\end{figure}

There are only 246,660 entities containing two or more addresses and these contains 1,516,130 addresses.
The number of non null elements in the rows of matrices $\mathbf{PreE}$ and $\mathbf{PostE}$ for the Entities Petri Net is reported in Figure \ref{fig_preE_CCDF} and \ref{fig_postE_CCDF} respectively. This corresponds the number of transactions where the entities are involved.
They clearly show a power-law distribution for transactions among the entities.

In Tab. \ref{tab_10Entity_1} we report the ten most used entities, 
found summing up the number of non zero elements in $\mathbf{PreE}$ rows
to that of non zero elements in $\mathbf{PostE}$ rows.
Their balances can be computed summing the balances of all addresses 
belonging to the corresponding entity and are owned by a single user. 

\begin{table*}[ht]
	\begin{center}

		{\begin{tabular}{ccccc} 

				\hline
				Entity number &  L pre  & L post & size & tags \\ \hline 
				95237  & 270,204  & 275,398 & 2 & deepbit.net \\ 
				2 & 102,186  & 283,973 &156,725& ilovethebtc\\ 
				37  & 51,228  & 147,712  &78,251& jmm5699 \\ 
				11 & 49,959  & 97,732 &10,37& - unknow - \\ 
				130& 20,857   & 58,350 &23,649& Instawallet\\ 
				66437 & 14,219   & 60,868 &13,289& Rai, Dread88 \\ 
				42 &  9,268  & 31,147 &10,561& Quip, iosp and other \\ 
				37598 & 8,923  & 31,004 &12,520& generalfault, safetyvest.com  \\ 
				220  & 11,133   & 27,487 &9,093& zephram  \\ 
				1503 & 9,044  & 29,400 &10,116&  folk.uio.no/vegardno \\ \hline

			\end{tabular}}

		\vspace{.2cm}
	\end{center}
	\caption{Summary of first 10 most active entities}
	\label{tab_10Entity_1}
\end{table*}

\begin{figure}
	\centering
	\includegraphics[width=7.5cm]{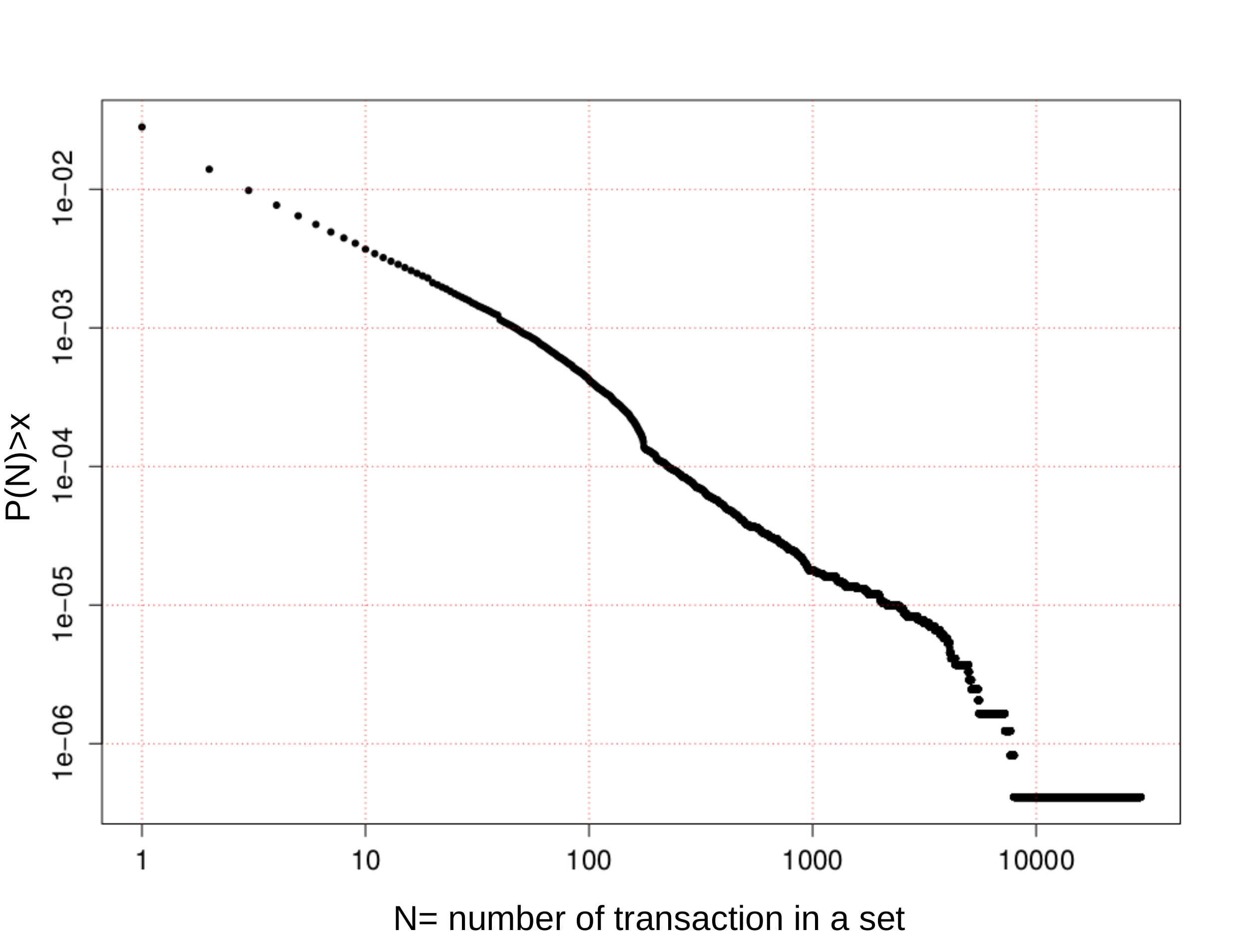} 
	\caption{CCDF of the size L of grouped transaction set for the Entity Petri Net net }
	\label{fig_transE_CCDF}

\end{figure}

Like for the Addresses Petri Net, we computed groups of repeated transitions for the Entities Petri Net. 
We found that about $22.6\%$ of transactions are a repetition of another one occurred among the same 
entities in input and in output.
This information allows to identify steady fluxes of bitcoins at the owners level. 
Figure \ref{fig_transE_CCDF} reports the CCDF for the sizes of these groups of repeated transactions.

\section{Discussion}
\label{sec:Discussion}
The Petri Net formalism can be natively used to infer many information and features about Bitcoin users and the Blockchain. 
We used the Petri Net model to gather together group of addresses (as entities or groups of disposable addresses) trying to associate an identity to each group. 
We also estimated how many users have been actually involved in the first three years and half of Bitcoin activity.

Analyzing the entities we found that 1,516,130 addresses are controlled by 246,660 owners at most.
With our model we were able to trace transactions chains whenever disposable addresses are involved. 
Each chain holds addresses belonging to one owner, but one owner may control more than one chain. 
So, according to our results, the 1,350,010 addresses involved are owned at most by 122,155 owners.
Using $pre$ and $post$ matrices we found that 609,295 addresses are used only as output of bitcoin transactions and are not involved in entities or chains. 
These three facts, enable us to estimate a threshold for the number of different Owners or users in the Bitcoin system.
We compute that there were 368,815 $engaged$ (or expert) owners that adopted disposable addresses practice or used two or more addresses in their operations. 
The addresses of such owners are involved in the $72,6\%$ of transactions. We suppose that the 609,295 addresses that appear only in output are used by some $engaged$ 
owners for the purpose of bitcoin depositing. Finally, the 255,045 remaining addresses are owned by occasional users.
Furthermore we tried to associate an identity to addresses showed in Tab. \ref{tab_10addressPreA} and to entities showed in Tab. \ref{tab_10Entity_1}.


Information about these addresses can be found on the Internet, in particular on blockchain explorer websites like \textit{blockchain.info}, or specialized forums like \textit{bitcointalk.org}.  
Some of theme are made more easily recognizable by attributing a \textit{tag}.
Take the case of the most used address we found, which appears 270,204 times as the input of a transaction. 
We were able to recognize that it belongs to a (now closed) Bitcoin \textit{pool} which was called \textit{DeepBit}.

Another example regards the most used entity in output that includes 156,725 addresses and regroup the $4.2\%$ of the total number of addresses. 
Searching on the Internet some of its addresses, we found out that who manages this entity has used varied tags, such as  \textit{ilovethebtc}, \textit{mikeo},
\textit{FredericBastiat}, \textit{edgeworth}, etc. 
\footnote{It is possible to find a portion of the addresses included in that entity, in the input section of a Bitcoin transaction, available on \textit{blockchain.info} and  reachable from this short link  \url{http://tinyurl.com/ilovethebtc}. Some of theme have a tag. } 
Several addresses in that entity have no tag. 
\\

From all the reported CCDFs it is clear that all the distributions are characterized by a strongly uneven amount of transactions across the addresses, either for pre and post transactions. 
This means that there are many addresses where the bitcoins are 
hardly exchanged, and few addresses where the rate of bitcoin 
exchange is particularly high. 
This analysis can be helpful for identifying addresses which are used by pool of miners. In fact, when miners join together in a pool to share computational facilities for mining operations, they need to define a common address where the mining rewards is accounted to. Then they need to redistribute the amount of gained bitcoins among all the pool users. As a consequence the 
address will be affected by a number of transitions in the corresponding Petri Net as large as the pool's size. 

Finally, since we analyzed a limited window of 180,000 blocks, the amount of transitions found in the matrices are also a signature of the average rate of Bitcoin transferred between different entities and such rate can be used to infer information on the organizations which can manage massive Bitcoin transfers.

\subsection{Advantages of the PT modeling}

In this section we discuss the motivations for preferring the PT formalism 
in modeling the bitcoin transactions on the blockchain (as well as other 
possible transactions) and describe the intrinsic advantages carried by 
this formalism. Part of this discussion will include proposals for further 
research. 

First, PT formalism allows for the "non determinism criteria" in the system's dynamics. 
Such criteria accounts for respecting the locality principle in the system's 
evolution. In other words, PT nets formalism natively 
includes independence between events generated by enabled transactions 
so that one enabled transaction can occur regardless the occurrence of 
other transactions for any given marking. Once an (or a set of) enabled 
transaction occurs he new marking has to be evaluated in order to 
understand which transactions are enabled in the new marking.

Such formalism perfectly fits into the blockchain transactions system 
where only transactions with non null UTXO are "enabled" and can occur, 
and their occurrence is independent from other transactions occurrence. 
A transaction occurrence is not deterministic and depends not only by 
the owner decision of sending bitcoin to another address, but also on the 
winning miners and on the probability that such transaction is included 
into the block validated by the hashing mechanism, which in turn has a 
different probability depending on the fee the owner accepts to pay.
In such model enabled transactions natively correspond to UTXO and 
the marking corresponds to the set of all UTXO determined by the last 
block validated. The validation of a new block, where transactions are 
included in an independent fashion, determines a new "marking" of the 
bitcoin PT net with a renewed set of UTXO. This enabling mechanism 
is hardly accounted for using a simple bipartite graph or a matrix 
representation for the bitcoin network and its transactions, 
even if many properties illustrated in this paper can be recovered 
by using such representations. The advantage of the PT nets 
formalism is that it natively includes such features.

The second aspect we discuss is related to simulation modeling 
which allows to analyze systems dynamics and which is a typical 
advantage provided by the PT nets formalism. 
Differently from bipartite graphs, which account for a static analysis, 
PT nets formalism includes systems dynamics and allows for 
non deterministic system's dynamics modeling. 
In fact in PT simultaneous transactions can occur provided 
they are not in conflict. 
Again this is a characterizing feature of bitcoin transactions 
dynamics where many non conflicting transfers of bitcoin between 
addresses can be included into the same validated block in the 
blockchain. Conflicting transactions, like for example double spending, 
are controlled and not allowed. Furthermore PT nets can 
include into the dynamics modeling priorities between transactions
and this can be used in a statistical modeling of the different
probabilities the bitcoin transactions have to occur depending 
on the fee the owner accepts to pay. 

A third feature natively included into the PT nets formalism
is the sequence of transactions: two transactions t1 and t2 are 
in a sequence if t1 precedes t2 with t1 enabled and t2 not enabled 
for a given marking, and when the occurrence of t1 enables t2 in the 
new marking. The bitcoin transactions network natively contains
sequences of transactions, like for example the sequences 
of UTXO generated into a single chain of disposable addresses 
monitored in our work and used for preserving bitcoin anonymity. 
Once again sequences of transactions can hardly accounted for 
using different representations, like bipartite graphs or 
matrices, without inserting ad hoc constraints into such 
representations. 

Another important advantage is that PT nets formalism includes 
the possibility to set state equations for the evolution 
dynamics. Given an initial marking the state equation 
allows the determination of the new marking according to 
the rules fixed for choosing the enabled transaction that 
effectively occur. The rules can be chosen with great 
freedom (respecting network constraints) and in particular 
a stochastic or probabilistic approach can be used in order
to simulate the evolution of the blockchain from a statistic 
point of view. For example, one of the future improvements  
the authors are presently working on is to collect 
statistics on the bitcoin fluxes between addresses paying 
attention to address clustered in entities, to addresses 
corresponding to exchanges and to addresses owned by miners 
pools, in order to assign transitions probabilities for 
bitcoin flux exchanges between such addresses to be 
used for choosing the enabled transactions to choose 
into the corresponding PT nets to make evolve its marking
using a statistical approach. This will provide a set 
of possible future marking, each with its own probability 
of occurrence, which will correspond to future states 
of the blockchain. Such statistical modeling can 
provides hints on which addresses are going to get 
richer with a given probability, which pools of miners 
are going to exploit the future mining and at which rate
and so on. The possibility of performing such a statistical 
simulation for the blockchain dynamics is straightforward 
within the PT nets model whilst is hardly accounted for 
using different approaches which are mainly static.

Last but not least, the formalism, through the use of pre and post matrices, 
allows to recover many different and independent results
following straightforwardly from standard computations over 
the pre and post matrices associated to the blockchain transaction network. 
For example, counting the number of rows and columns of matrices $\mathbf{PreA}$ or $\mathbf{PostA}$ 
it is straightforward to find the number of addresses and transactions, or we can find bitcoin addresses never used to spend
looking at addresses with only zeros in $\mathbf{PreA}$ rows and at least one non null element in $\mathbf{PostA}$ rows, 
or we can recover $disposable\ addresses$ looking at transitions that correspond to columns of $\mathbf{PreA}$ having only one non zero 
element and to columns of $\mathbf{PostA}$ having two non zero elements but in different rows.

\section{Conclusions}
\label{sec:Conlcusions}
In this paper we introduced a novel approach, based on a Petri Net model to parse the Blockchain. Our purpose was to define a single useful model in which all main information about transactions and addresses are represented. 
Collecting the first 180 thousand blocks, we were able to associate a place for each address and a transition for each bitcoin transaction. Our Petri net includes \textit{pre} and \textit{post-incidence} matrices where all links between addresses and transactions are modeled.


\Andrea{ESPANDERE} We are aware about the limitation of computational problem of a matrix approach. The portion of Blockchain which we chosen was processed without specific hardware resources. Anyway, the current size of the Blockchain (over 480,000 blocks and the total number of transaction is over 240 million) could not allow us to handle easily all the blocks information.

\Andrea{Spostato per evidensizare risultati}
 
However, by using this model, we were able to pick out significant and  original results. This formalism has proven powerful methodology for performing many kinds of measurements and analysis. Analyzing the number of pre and post arcs, we had proof of the presence of power-law like distributions. We made use of both incidence matrices for determining all transactions chains, identifying a typical \textit{disposable addresses} usage by Bitcoin users. By measuring the chains' lengths, we found again 
power-law like distributions.
We were also able to determine that some transactions involve the same group of address in input and in output. We gathered these transaction in sets and the size of such sets follow again a power-law like distribution. 
By reading information of \textit{pre-incidence} matrix, we were able to identify the entities and we built the Entities Petri Net, repeating on such Petri net all the measures done for the Addresses Petri Net.

Furthermore, despite the current Blockchain size is about two order of magnitude greater than the size of the portion that we have studied, our approach can be adopted to study a specific portion of the Blockchain. For example starting from a specific set of addresses which we want to investigate and analyze. In fact, it is always possible to build the addresses Petri Net containing only the part of blockchain of interest.

On the basis of all the obtained results, we believe that our model can be used for studying a large set of other issues related to other systems based on Blockchain technology, such as Ethereum. Today, Ethereum attracts increasing attention and will be one of our future research topic.

\newpage

\balance




\appendix
\section{Chains of disposable addresses}
\label{apx:chains}
We model the Blockchain as a Petri net, a bipartite oriented graph $N$,  defined as $N=(P_\alpha,T,Pre,Post)$, where $P_\alpha$ is the set of the \textit{places (addresses)}, $ T$ is the set of the  \textit{transitions (transactions)}, $Pre$ is the \textit{Pre-incidence } matrix and $Post$ is the \textit{Post-incidence} matrix. The element $ij$ in the Pre matrix defines  how many times the address $i$ is in the input section of the transaction $j$, instead, in the Post matrix, it defines how many times the address $i$ is  in the output section of the transaction $j$. After having built of the Petri net $N$, we focus our attention on the chains of disposable addresses, 
and hence on the transactions having only one address, $\alpha_a$, in the input section and only two addresses, $\alpha_b$ and $\alpha_c$, in the output section. In more detail, the address $\alpha_a$ in the input section is used by a user $u_1$ to send bitcoins to one of the addresses in the output section, $\alpha_b$, belonging to a user $u_2$. The other  address, $\alpha_c$, in the output section is created by the user $u_1$ to collect the change.
We created the set of potentially disposable addresses $A_d$, starting from the set $A$ of the addresses $\alpha$ and from the set  $\Theta$ of the transactions $\theta$ in the Blockchain.

Let $\Theta_d$ be the set of transaction $\theta_d$ such that:
\[\Theta_d \subseteq \Theta = \{ \theta_d : \vert IN(\theta_d)\vert = 1,  \vert OUT(\theta_d)\vert = 2,\\
\vspace{-.2cm}
IN(\theta_d) \in  A_d,\] \[\exists \ \alpha \in OUT(\theta_d) : \alpha   \in  A_d, \forall \theta_d \in \Theta_d \}.\] 

In order to build a chain, for each $\theta_d$ we need to know the previous transactions $\theta_{dp} = PREV(\theta_d)$.
Using $Pre$ and $Post$ matrices, it is very easy to look for these previous transactions. 
We call $\Theta_{ds} \subseteq \Theta_d$ the set of transaction $\theta_{ds}$ that could be considered the starting point of a chain because it does not have a previous transaction inside $\Theta_d$ . We denote with $\alpha_{ds}$ the address in input to a transaction $\theta_{ds}$.
Finally, we call $NEXT(\theta_d)$ the transaction $\theta_{d'}$ which has, in the input section, the disposable address that is contained in the output section of the transaction $\theta_d$.
To find the chains $c$ of disposable addresses, we defined and implemented the following algorithm: 

\begin{enumerate}
	\item Let $C = \emptyset$ be a set of empty chains, $c$,
	\item for each $\theta_{ds}  \in \Theta_{ds}$:
	\begin{enumerate}   
		\item take an empty chain, $c$,
		\item insert $\theta_{ds}$ in $c$
	\end{enumerate}
	endfor
	\item for each $c \in C$
	\begin{enumerate}
		\item take the last element inserted in $c$, $\theta_d$,
		\item while $\exists \ \theta_{d'} = NEXT(\theta_d)$
		\begin{enumerate}
			\item insert $\theta_{d'}$ in c\\
			endwhile
		\end{enumerate}
	\end{enumerate}
	endfor
\end{enumerate}

The algorithm returns a set $C$ of chains $c$. Each chain $c$ contains the transactions ordered by execution order.


\end{document}